\documentclass[num-refs]{wiley-article-modified}

\usepackage{graphicx}
\usepackage[space]{grffile}
\usepackage{latexsym}
\usepackage{textcomp}
\usepackage{longtable}
\usepackage{tabulary}
\usepackage{booktabs,array,multirow}
\usepackage{amsfonts,amsmath,amssymb}
\usepackage{natbib}
\usepackage{url} 
\usepackage{hyperref} 
\hypersetup{colorlinks=false,pdfborder={0 0 0}}
\usepackage{etoolbox}
\newif\iflatexml\latexmlfalse

\AtBeginDocument{\DeclareGraphicsExtensions{.pdf,.PDF,.eps,.EPS,.png,.PNG,.tif,.TIF,.jpg,.JPG,.jpeg,.JPEG}}

\usepackage[utf8]{inputenc}
\usepackage[ngerman,english]{babel}

\usepackage{siunitx}

\newcommand{\MSE}{\text{MSE}}
\newcommand{\ESS}{\widehat{\text{ESS}}}
\newcommand{\CV}{\text{CV}}

\newcommand{\bI}{{\bf I}}
\newcommand{\x}{{\bf x}}

\newcommand{\w}{{\bf w}}

\newcommand{\X}{{\bf X}}

\newcommand{\cblue}{\textcolor{black}}

 \newcommand{\cgreen}{\textcolor{black}}
\newcommand{\cmag}{\textcolor{black}}

\newcommand{\ccyan}{\textcolor{black}}

\newcommand{\Real}{\mathbb{R}}  
\newcommand{\normalized}{\widetilde}

\newcommand{\E}{\mathrm{E}}
\newcommand{\Var}{\mathrm{Var}}
\newcommand{\Cov}{\mathrm{Cov}}

\iflatexml

\else

\paperfield{}
\abbrevs{ESS, effective sampling size; IS, importance sampling; MC, Monte Carlo; MCMC, Markov chain Monte Carlo.}
\corraddress{School of Mathematics \\
University of Edinburgh\\
James Clerk Maxwell Building\\
The King's Buildings\\
Peter Guthrie Tait Road\\
Edinburgh
EH9 3FD
United Kingdom}
\corremail{victor.elvira@ed.ac.uk}
\presentadd{}
\fundinginfo{}
\fi

\title{Rethinking the Effective Sample Size}

\author[1,2]{V\'ictor Elvira}
\author[3]{Luca Martino}
\author[4]{Christian P. Robert}

\affil[1]{University of Edinburgh, United Kingdom}
\affil[2]{The Alan Turing Institute, United Kingdom}
\affil[3]{King Juan Carlos University  of Madrid, Spain}
\affil[4]{Universit\'e Paris Dauphine, France}

\runningauthor{V\'ictor Elvira}

\begin{document}

\maketitle
\selectlanguage{english}
\begin{abstract}
The effective sample size (ESS) is widely used in sample-based simulation
methods for assessing the quality of a Monte Carlo approximation of a given
distribution and of related integrals. In this paper, we revisit the
approximation of the ESS in the specific context of importance sampling (IS). The
derivation of this approximation, that we will denote as $\ESS$, is 
partially available in \cite{Kong92}. This approximation has been widely used
in the last 25 years due to its simplicity as a practical rule of thumb in a
wide variety of importance sampling methods. However, we show that the multiple
assumptions and approximations in the derivation of $\ESS$, makes it difficult
to be considered even as a reasonable approximation of the ESS. {We extend the
discussion of the $\ESS$ in the multiple importance sampling (MIS) setting, we display
numerical examples, and we discuss several avenues for developing alternative metrics.} This paper does not cover the use of ESS for MCMC algorithms.

\end{abstract}%

\section{Introduction}
 
A usual problem in statistics, signal processing, and machine learning consists
in approximating an intractable integral \cblue{with respect to (w.r.t.)} a targeted distribution. In the
simplest Monte Carlo algorithm, the integral is approximated by random samples
simulated from the targeted distribution itself \cite[e.g.,][Section
3.2]{Robert04}. However, in many problems it is not possible to simulate from
the targeted distribution or it is not efficient to do so. Importance sampling
(IS) is a well-established Monte Carlo technique where samples are drawn from a
different distribution (called \emph{proposal} distribution), and an importance
weight is assigned to each sample to take into account the mismatch between
proposal and targeted distributions \citep[e.g.,][Section 3.3]{Robert04}.

Monte Carlo methods are usually evaluated by the mean square error (MSE) of the estimators that approximate the integral of interest. In many cases, these
estimators are unbiased (as, e.g., for the plain Monte Carlo estimator) or the
bias quickly vanishes when the number of samples increases (as, e.g., for the
self-normalized IS (SNIS) estimator). Therefore, Monte Carlo estimators are
most often characterized and compared through their variance \citep[Chapter
9]{owen2013montecarlo}. 
\cblue{This is the case of the effective sample size (ESS), which involves the ratio of the variances of the plain Monte Carlo and SNIS estimators.} 
However, computing these variances usually requires solving
integrals of a similar complexity as the (intractable) integral at the source of the analysis.
Hence, the problem to solve is two-fold. First, the integral of interest is
intractable. Second, characterizing the method that approximates the integral best
is also impossible, because of a second intractable integral. An alternative
that bypasses this deadlock consists in a relative comparison of the performances
of different Monte Carlo estimators. 
In other words, instead of characterizing in absolute terms a specific estimator, one
can compare two or more Monte Carlo estimators. However, a na{\"\i}ve  comparison of these via a second Monte Carlo experiment is almost always too costly to be considered.

Conceptually, the ESS can be and often is interpreted as the number of particles/samples that would need to be
drawn from the target (\emph{direct} or \emph{raw} Monte Carlo) in order to yield the same variance as
that from the self-normalized IS estimator under study  \citep{Kong92,Liu95}. 
However, this relative measure is itself impossible to obtain in a closed form in most cases. 
Indeed, it is proportional to the ratio of the variances of the two
aforementioned estimators, variances that are impossible to compute. 
For this reason, an approximation of the ESS has been used instead for the last
two decades in the Monte Carlo literature. Its simple expression is given by
\begin{equation}
\ESS = \frac{1}{\sum_{n=1}^N \bar w_n^2},
\end{equation}
where $\bar w_n$ represent the normalized importance weights associated with the $N$ samples.
\cblue{The approximation from ESS to $\ESS$ is described in the technical report of \cite{Kong92}, although the final part is not fully explicit (see more details in next section).}
\cblue{Since the technical report \cite{Kong92} and the follow-up paper
\cite{Kong94}, the $\ESS$ metric has been widely
used (and sometimes misused) as an approximate relative measure of performance between the self-normalized IS
estimator and the vanilla Monte Carlo estimators.\footnote{\cblue{The technical report \cite{Kong92} can be alternatively found in \url{https://victorelvira.github.io/papers/kong92.pdf}.}}  
We highlight that the $\ESS$ was derived in 1992 and has been applied in IS-based methodology that did not exist at that time (e.g., particle filters were proposed later in \cite{Gordon93} and also the first adaptive IS samplers \cite{Cappe04}).
While it is hard to
identify the number of publications where the ESS is used, the authors estimate
it to be in the thousands.
It is therefore of clear interest to understand the
derivation of $\ESS$, including its approximations, assumptions, and resulting
limitations.}

\cblue{In this paper, we first review in detail the derivation of $\ESS$, revisiting \cite{Kong92}. We clarify all steps, exposing all assumptions and approximations that are
introduced with the goal of simplifying intractable terms and producing a
computable expression. Then, we summarize the drawbacks of these approximations, focusing on situations that lead to serious problems and misleading conclusions. The main goal of this paper is to raise consciousness around the $\ESS$, so practitioners are aware of its flaws and can be careful when they decide to use it.} 
Note that in this work, we focus on the definition of ESS in IS, and not in the ESS used in Markov Chain Monte Carlo (MCMC) which was first defined in \cite{sokal1997monte}.

The rest of the paper is organized as follows. In Section \ref{sec_is}, we briefly describe IS and the ESS. In Section \ref{sec_approx}, we completely derive the approximated  $\ESS$ from ESS, and the assumptions, approximations, and consequences are clearly listed. \cblue{We analyze other derivations and connections of the $\ESS$ in Section \ref{sec_beyond}, along with its extension in MIS. We also review existing alternative metrics, and we include three numerical examples. In Section \ref{sec_new}, we discuss new directions to develop novel metrics with the purpose of overcoming undesired limitations of the $\ESS$.} We conclude the paper with a discussion and some possible future lines in Section \ref{sec_conclusions}.
 
\section{Effective Sample Size in Importance Sampling}
\label{sec_is}
\subsection{Importance sampling}

Let us consider the problem of estimating the integral 
\begin{equation}
I = \int h(\x) \normalized \pi(\x)d\x,
\label{eq_problem}
\end{equation}
where $\normalized \pi$ is a targeted pdf that can be evaluated up to a normalizing constant, and $h$ is an integrable function w.r.t. $\normalized \pi$. We thus assume that $\normalized \pi$ is known to an unknown normalizing constant $Z$, 
$$\normalized \pi(\x) = \frac{\pi(\x)}{Z},$$
where obviously $Z = \int \pi(\x)d\x$,
meaning that only the non-negative function $\pi$ can be evaluated.
In the simplest version of IS, $\normalized \pi(x)$ is approximated with a set of $N$ weighted samples as follows. First, $N$ samples \cblue{$\X_1,...,\X_N$}  are drawn from a proposal density $q(\x)$. Second, each sample is assigned with an importance weights as $W_n = \frac{\pi(\X_n)}{q(\X_n)}$. 
\cblue{The unbiased estimator of $I$, {often called \emph{unnormalized IS} (UIS) estimator}, is generally used if the normalizing constant $Z$ is known \cite{elvira2021advances}. 
Moreover, in} the general case where $Z$ is unknown, the self-normalized IS (SNIS) estimator can always be used:
\begin{equation}
\widetilde I = \sum_{n=1}^{N} \overline W_n h(\X_n), \qquad \X_n \sim q(\x),
\label{eq_SNIS}   
\end{equation} 
where $\overline W_n = \frac{W_n}{\sum_{i=1}^N W_i} $ are the normalized weights.

\cblue{ 
In the case of the UIS estimator, its variance is proportional to the $\chi^2$-divergence between a normalized version of the function $\normalized\pi(\x)|h(\x)|$ and the proposal $q(\x)$ (see e.g., \cite{Robert04, kahn1953methods}).  
The optimal proposal for the SNIS estimator, $\widetilde I$, is  $q^*(\x)\propto |h(\x) - I|  \pi(\x)$ \citep{hesterberg1988advances,owen2013montecarlo,elvira2021advances}. A possible interpretation is that the optimal proposal in SNIS trades-off minimizing both the aforementioned $\chi^2$-divergence and also the $\chi^2$-divergence between the target pdf (i.e., without the function $h$) and the proposal.}
It is therefore a legitimate question to examine the efficiency of the set of weighted samples simulated from a specific proposal $q$.
\subsection{The effective sample size}
 
The effective sample size (ESS) is an indicator of the worth of the IS estimator defined as the number of samples simulated from the target pdf $\normalized \pi$ that would provide an estimator with a \emph{performance} equal to the performance of the IS estimator based on $N$ samples (drawn from the proposal pdf). 
More precisely, if it were possible to simulate samples from the target $\normalized \pi$, the integral $I$ could also be approximated by
\begin{equation}
\bar I = \frac{1}{N} \sum_{n=1}^{N} h(\X_n), \qquad \X_n \sim \normalized \pi(\x).
\label{eq_estimator_target}
\end{equation} 
In \cite{Kong92}, the ESS is defined as the ratio between the variance estimator with $N$ samples from the target, and the variance of the IS estimator. In particular, the ESS of the self-normalized estimator is given by
\begin{equation}
\text{ESS} =N\frac{\Var[\bar I]}{\Var[\widetilde I]},
\label{eq_var_var}
\end{equation}
which is approximated in \cite{Kong92} and applied in \cite{Kong94} for a missing data problem. Note that \text{ESS} depends on a specific integrand $h(\cdot)$, which means that the same set of weighted samples can be suitable for a specific function $h_1(\cdot)$ but disastrous for another function $h_2(\cdot)$.

When devising IS-based methods, an assessment of the ESS is of considerable interest, since
it provides a direct measure of the efficiency of a potential method.
Unfortunately, obtaining analytically this ESS is not possible in most of the
scenarios where IS methods are applied; similar integrals to that of Eq.
\eqref{eq_problem} need be computed, e.g., the variance of $\bar I$ is given by

\begin{eqnarray}
\Var_{\normalized \pi}[\bar I] &=& \frac{1}{N} \left(  \E_{\normalized \pi}[H^2]  - \left(\E_{\normalized \pi}[H] \right)^2 \right),  
\label{eq_var_target_sampling}
\end{eqnarray}
where we denote $H\equiv h(\X_n)$ to alleviate notations.\footnote{Note that,
here, all samples are i.i.d. so we can drop the subindex $n$ from the summation.} Unfortunately,
neither $\E_{\normalized \pi}[H^2] = \int h^2(\x) \normalized \pi(\x)d\x$ nor
$\E_{\normalized \pi}[H] =  \int h(\x) \normalized \pi(\x)d\x $ can be computed
in most problems of interest. 
Moreover, the exact variance of the SNIS estimator, $\widetilde I$, has almost
never a closed form. Due to the intractability of the ESS, the ratio in Eq.
\eqref{eq_var_var} is most often approximated in the Monte Carlo literature by the expression
\begin{equation}
\ESS = \frac{1}{\sum_{n=1}^N \bar w_n^2}.
\label{eq_rule_of_thumb}
\end{equation}
In the following we describe the path leading from \eqref{eq_var_var} to the approximation in \eqref{eq_rule_of_thumb}.

\section{Approximating the effective sample size}
\label{sec_approx}

In the
following, we first detail all the steps in the derivation of the approximation
$\ESS$. Then, we summarize the different assumptions and approximations behind
$\ESS$. Finally, we analyze the consequences of these assumptions and approximations.

\subsection{From ESS to $\ESS$}
\label{sec_derivation}

\cblue{First, note that $\widetilde I$ is a ratio of two dependent r.v.'s and a biased estimator of $I$.    
The bias of $\widetilde I$ can be considered negligible w.r.t. its variance when $N$ grows, since the bias is typically $\mathcal{O}(N^{-1})$, so that the bias squared is 
$\mathcal{O}(N^{-2})$, while the variance is 
$\mathcal{O}(N^{-1})$ \cite{Robert04}. Nevertheless, for low or intermediate $N$, the bias might be far from negligible (see, e.g., \cite{owen2013montecarlo}).} A more relevant expression for a relative comparison of performance might be
\begin{equation}
\text{ESS}^* = N\frac{\Var[\bar I]}{\MSE[\widetilde I]}. %
\label{eq_ess_mse}
\end{equation} 
Otherwise, using IS with $N$ simulations does not exactly produce the same MSE as using direct simulation with ESS samples from the target.
Therefore, a first implicit approximation involved in the handling of ESS is
\begin{eqnarray}
\MSE_q[\widetilde I]   \approx {\Var_q[\widetilde I]}.
\end{eqnarray}

\cblue{Note that $\widetilde I$ in Eq. \eqref{eq_SNIS} is the ratio between $N$ realizations of the r.v.'s $WH$ and $W$, and thus both numerator and denominator are dependent. 
Then, the variance of $\widetilde I$ in Eq. \eqref{eq_SNIS} cannot be computed in a closed form and must be approximated.
It is possible to use the delta method, which is a second order Taylor expansion used to approximate expectations of functions involving two dependent r.v.'s \cite{rice2006mathematical}. In Appendix \ref{appendix_delta_1}, we provide the full derivation of the delta method for generic functions, and here we apply the resulting Eq. \eqref{delta_var_2} for the case where the function is a ratio, $f(T,Y)=\frac{T}{Y}$, with $T=WH$ and $Y=W$. In this case, the expansion is done around the expectation of each r.v., i.e.,  $(T_0,Y_0) = (IZ,Z)$. The first derivatives are $f'_T(T,Y) = \frac{1}{Y} = \frac{1}{W} $ and $f'_Y(T,Y) = -\frac{T}{Y^2} = -\frac{H}{W}$. This yields} 
 
\begin{align}
{\Var_q[\widetilde I]}&\approx \frac{1}{N} \left( \left(\frac{\E_q[HW]}{\E_q[W]}\right)^2\Var_q[W] + \Var_q[HW] -2 \left(\frac{\E_q[HW]}{\E_q[W]}\right)\Cov_q[W,HW]  \right) \nonumber \\
&= \frac{1}{N} \left( I^2 \Var_q[W] + \cgreen{\Var_q[HW]} -2I\cmag{\Cov_q[W,HW]} \right),
\label{eq_delta_1}
\end{align}
where we use the identities $\E_q[HW] = ZI$ and $\E_q[W]=Z$.  
\cblue{In the following, we assume without loss of generality that the target is normalized,
i.e., $Z=1$, as it is done in \cite{Kong92}. It is easy to show that the derivation holds for any $Z$ by realizing that the normalized weights in $\widetilde I$ do not depend on $Z$.} 
Let us expand the expressions 
\begin{eqnarray}
\cmag{\Cov_q[W,HW] }&=& \E_q[HW^2]-\E_q[W]\E_q[WH] \nonumber\\
&=& \E_q[HW^2] - I  \nonumber \\
&=& \E_{\tilde \pi}[HW] - I \nonumber \\
&=& \Cov_{\tilde \pi}[W,H] + I \E_{\tilde \pi}[W] - I,
\label{eq_cov_w_hw}
\end{eqnarray}
and
\begin{eqnarray}
\cgreen{\Var_q[HW]} &=& \E_q[W^2H^2]-\E_q^2[WH] \nonumber \\
&=&  \ccyan{\E_{\tilde \pi}[WH^2]}- I^2.
\label{eq_var_hw}
\end{eqnarray}

 \cblue{The delta method is again applied to approximate the expectation ${\E_{\tilde \pi}[WH^2]}$ using the first two moments of $W$ and $H$, i.e., as a second order Taylor approximation of this expectation. In this case, we apply the generic approximation for expectations in  Eq. \eqref{delta_expectation} in Appendix \ref{appendix_delta_1}, particularizing for function $f(T,Y)=TY^2$, where $T=W$ and $Y=W$:}

\begin{eqnarray}
{\E_{\tilde \pi}[WH^2]} &\approx& \E_{\tilde \pi}[W]\E^2_{\tilde \pi}[H]+\frac{1}{2}\Var_{\tilde \pi}[H](2\E_{\tilde \pi}[W])+\Cov_{\tilde \pi}[W,H](2\E_{\tilde \pi}[H]) \nonumber \\
&=& I^2 \E_{\tilde \pi}[W]+\Var_{\tilde \pi}[H]\E_{\tilde \pi}[W]+2I\Cov_{\tilde \pi}[W,H].
\label{eq_delta_2}
\end{eqnarray}
\cblue{The remainder term in this approximation is the expectation of a polynomial order three, namely $\E_{\tilde \pi}[(W- \E_{\tilde \pi}[W])(H-\E_{\tilde \pi}[H])^2] = \E_{\tilde \pi}[(W-\E_{\tilde \pi}[W])(H-I)^2]$. This is easy to show by generalizing Eq. $\eqref{taylor_order_2}$ in Appendix \ref{appendix_delta_1} to a third order Taylor expansion.}
 
Then, plugging Eqs. \eqref{eq_cov_w_hw}, \eqref{eq_var_hw}, and \eqref{eq_delta_2} in Eq. \eqref{eq_delta_1}, we obtain
\begin{eqnarray}
{\Var_q[\widetilde I]} &\approx& \frac{1}{N} \left( {\Var_{\tilde \pi}[H]}\E_{\tilde \pi}[W] + I^2 \left(1+\Var_q[W] - \E_{\tilde \pi}[W] \right) \right) \nonumber\\
&=& \Var_{\tilde \pi}[\bar I] \left(1+ \Var_q[W] \right),
\label{eq_aux1}
\end{eqnarray}
 
where in the last equality, we use $\frac{1}{N} { \Var_{\tilde \pi}[H]} = \Var_{\tilde \pi}[\bar I]$ (see estimator of Eq. \eqref{eq_estimator_target}) and $\E_{\tilde \pi}[W] = \E_q[W^2]=\Var_q[W]+1$ \cblue{(we recall we assume $\E_q[W]=Z=1$ without loss of generality)}.
Therefore, using Eq. \eqref{eq_aux1}, the ESS is approximated as
\begin{eqnarray}
\text{ESS} = N\frac{\Var_{\tilde \pi}[\bar I]}{\MSE_q[\widetilde I]} \approx N\frac{\Var_{\tilde \pi}[\bar I]}{\Var_q[\widetilde I]} \approx \frac{N}{1+\Var_q[W]}.
\label{eq_ess_1}
\end{eqnarray}
which implies that ESS is always {\em less than} $N$, a drawback discussed
below. Note that the derivation of \cite{Kong92} ends at this point. In
\cite[Section 4.1]{Kong94}, the authors state that \emph{``although there is no
guarantee that the remainder term in this approximation is ignorable, what is
nice about Eq. \eqref{eq_ess_1} is that it does not involve $h$''.  }%
First, the fact that the expression does not depend on $h$
indicates that it is necessarily a loose approximation, since the quality of
the IS approximation depends on the mismatch of the proposal $q(\x)$ w.r.t.
$|h(\x)|\pi(\x)$ \citep[e.g.,][]{Robert04, kahn1953methods} \cblue{(see also the example in Section \ref{sec_rare_toy})}. 

Second, other approximations can be derived from the simulated sample, including
unbiased estimates of $\E_q[W^2H^2]$ and $\E_q^2[WH]$, obtained at the same
cost as the original estimates, as well as bootstrap versions. At last, we
point out the most detrimental aspect of \eqref{eq_ess_1}, namely that it
always increases the variance from the i.i.d.~simulation reference, when it is
well-known that there always exist IS versions that on the contrary decrease
the variance (see Section \ref{sec_conseq} for more details).

\cblue{The further approximation from Eq. \eqref{eq_ess_1} to $\ESS =
{1}\big/{\sum_{n=1}^N \bar w_n^2}$ does not appear in \cite{Kong92}. Note also that the derivation assumes twice a normalized target for the approximation of Eq.
\eqref{eq_ess_1}. 
This is without loss of generality, since the SNIS estimator in \eqref{eq_SNIS} can be rewritten as}
 
\begin{equation} 
\widetilde I = \sum_{n=1}^{N} \frac{W_n}{\sum_{i=1}^N W_i} h(\X_n), \qquad \X_n \sim q(\x).
\end{equation}
Then, it becomes evident that the estimator remains the same (and hence its variance) if all weights are multiplied by a constant. Hence, one can assume in the previous formulation that the weights are $\widetilde W_n = \frac{W_n}{Z}$ instead $W_n$, which yields $\E_q[\widetilde W]=1$ true.
In that case, \eqref{eq_aux1} turns
\begin{eqnarray}
\text{ESS} \approx \frac{N}{1+\Var_q[\widetilde W]}.
\label{eq_ess_1b}
\end{eqnarray}
Then, if $Z$ is known, Eq. \eqref{eq_ess_1b} can be adapted as
\begin{eqnarray}
\text{ESS}  &\approx& \frac{N}{1+\frac{\Var_q[W]}{Z^2}}\\
&=&  N\frac{Z^2}{\E_q[W^2]},
\label{eq_ess_2}
\end{eqnarray}
where we have used $\Var_q[W] = \E_q[W^2] -\E_q^2[W]  = \E_q[W^2] - Z^2$.\footnote{The authors have unsuccessfully tried to find a published article with the remaining derivation from \eqref{eq_ess_1} to \eqref{eq_ess_3}. The first and only reference from this final part can be found in the blog post of one of the authors \citep{RobertBlogESS}.} In practical scenarios, neither $Z$ or $\E_q[W^2]$ is known. \cblue{A particle approximation of the expectations can be used to produce the estimator}
\begin{eqnarray}
\ESS  &=&  N\frac{\left(  \frac{1}{N} \sum_{n=1}^N  w_n \right)^2}{  \frac{1}{N} \sum_{n=1}^N  w_n^2  }  \label{eq_ess_3a} \\
&=&  \frac{1}{\sum_{n=1}^N \bar w_n^2},
\label{eq_ess_3}
\end{eqnarray}
where we use $Z \approx \frac{1}{N} \sum_{n=1}^N w_n$ and $\E_q[W^2]   \approx \frac{1}{N}\sum_{n=1}^N w_n^2$. \cblue{We remark that, although increasing $N$  improves the quality of the particle approximation in the last step, this is not a guarantee of $\ESS$ getting closer to ESS, as we show in the example of Section \ref{eq_ex1}.}

\subsection{Summary of assumptions and approximations}
\label{sec_ass_and_app}
The question at this stage is whether or not $\ESS$ is a good mathematical
approximation of the effective sample size, but also whether or not it keeps
some of its desirable properties. Let us summarize the assumptions and approximations behind
the derivation of  $\ESS$:  
\begin{enumerate}
\item The ESS is conceptually defined as the ratio between the performance of
two estimators, the direct Monte Carlo estimator, ${\bar I}$, where samples are
drawn from the target $\bar \pi$, and the SNIS estimator ${\widetilde I}$. The
expression $\text{ESS}=N{\Var_{\tilde \pi}[{\widehat
I}]}\big/{\Var_q[{\widetilde I}]}$  in Eq. \eqref{eq_ess_1} does not take in account
the bias of ${\widetilde I}$  (which can be significant for small $N$).
Then, the ratio of variances overestimates the theoretical value
ESS$=N{\Var_{\tilde \pi}\left[{\bar I}\right]}\big/{\MSE_q\left[{\widetilde
I}\right]}$, especially for low $N$. This can be particularly dangerous since
the approximated ESS can report a good performance in a scenario with large
bias. \cblue{However, the error of this first approximation vanishes as $N$ grows.}
\item In the derivation of \cite{Kong92}, all the samples are considered to be
i.i.d. from a unique proposal $q$, i.e., $\X_n \sim q(\x)$, for $n=1,...,N$.
Nevertheless,  $\ESS$ is used mostly in algorithms with multiple proposals, as
in adaptive importance sampling (AIS) or sequential Monte Carlo methods (SMC)
methods
\citep{Djuric03,Doucet08tut,Chen03tecRep,Cappe04,SMC01,elvira2017improving,Gordon93,martino2017layered,elvira2019langevin}. {It is worth noting that controlling the $\ESS$ has been proved to guarantee the long term stability of sequential Monte Carlo methods under certain assumptions \cite{whiteley2016role} (see Section \ref{rho_interpr}).} 
Extensions to dependent samples as in MCMC are even more delicate.
\item A first delta method is applied in order to approximate
$\Var_q[\widetilde I]$ in Eq. \eqref{eq_delta_1}. Then, the delta method is applied a second time to approximate
$\ccyan{\E_{\tilde \pi}[WH^2]}$ in Eq. \eqref{eq_delta_2}. 
\end{enumerate}

\subsection{Some {(undesirable)} consequences}
\label{sec_conseq}
\cblue{As a consequence of the unrealistic assumptions and approximations (mostly the Taylor expansions in the delta methods)} in the
derivation of $\ESS$, this approximated ESS exhibits some flawed features that
parallel the flaws in its derivation.
\begin{enumerate}
\item \cblue{The approximation is bounded as}
$$1 \leq \ESS \leq N.$$
\cblue{We provide a proof in Section \ref{sec_euclidean_distance}. This leads to two undesired consequences:}
\begin{itemize}
\item Even when the sample approximation to the integral $I$ is very poor, we
have that $\ESS \geq 1$. Note that the lower bound of the (exact) ESS is $0$,
corresponding to the case when the IS estimator has infinity variance and is
completely inappropriate. That $\ESS$ cannot detect infinite variance estimator
appears to us as a major defect of this indicator.
\item The bound of  $\ESS\leq N$ is equally limiting. In many settings,
$\Var_q(\widetilde I) \leq \Var_{\tilde \pi}[\bar I]$ for an adequate choice of the proposal $q$.
For instance, in connection with several advanced IS techniques such as PMC \citep{Cappe04}, AIS
methods \citep{bugallo2017adaptive}, a smart choice of
MIS schemes \citep{elvira2019generalized}, or variance reduction techniques
\citep[Section 8]{owen2013montecarlo}, importance sampling estimates can significantly lower the variance of the i.i.d.~solution. In such scenarios, ESS is {\em larger}
than $N$, but $\ESS$ cannot capture this variance reduction. Note that in the
extreme case where the theoretically optimal proposal is used, {the UIS estimator can achieve
zero variance, but also the SNIS estimator can largely reduce variance w.r.t. \cblue{standard Monte Carlo}}. Hence, a good approximation of ESS should never be upper bounded.
\end{itemize}
\item $\ESS$ does not depend on the function $h$. While this is an advantage
for practical reasons, the ESS conveys the efficiency or lack thereof of the
approximation of the moment in $h$ of $\normalized \pi$, but this dependence is
completely lost in the approximation. For the same target distribution, a
particle approximation can provide a low-variance estimator of some moment
$h_1$ and be disastrous for another moment $h_2$. {See Section \ref{sec_rare_toy} for a toy example of estimation of rare events.}  
\item $\ESS$ does not depend on the set samples $\{\X_n \}_{n=1}^N$, but only in their associated (normalized) weights $\{\bar w_n \}_{n=1}^N$. \cblue{In other words, the diversity of the sample cannot be directly evaluated through the corresponding weights, which are a mapping from the dimension of $X$ to dimension one. While this might seem to be a feature, $\ESS$ clearly lacks the ability to assess the diversity of the sample approximation, which is key to evaluate its worth. It is well-known that having diversity in the samples is essential to improve the performance of the estimators, e.g., negative correlation is induced to reduce their variance in antithetic and stratified sampling \cite[Chapter 8]{owen2013montecarlo}.} {This problem is also evident when resampling steps are needed, e.g., in particle filtering and other iterative or adaptive IS-based methods \citep{Douc05,li2015resampling}.} Note that, despite a potential inadequacy of the sampling approximation, the ESS is always equal to $N$ after a resampling step since all the weights are then set equal, even if the same particle was replicated $N$ times (potentially, in the worst scenarios, in an area of the target with low mass or where $h$ can be zero).
\end{enumerate}

\subsection{\cblue{Examples}}
\cblue{In the following, we provide two examples that allow us to evaluate the behavior of the $\ESS$, and a third example with multiple proposals is provided in Section \ref{sec_MIS}. In the first example, we study how the disparity between ESS and $\ESS$ grows when the mismatch between target and proposal densities is increased. The second example focuses on the estimation of rare events. More precisely, $h$ is an indicator function  that takes value one in a subset of the support where the target density has low probability mass. This example highlights the disadvantage of not including function $h$ in $\ESS$. }

\subsubsection{\cblue{ESS and $\ESS$ depending on the mismatch between target and proposal}}
\label{eq_ex1}

\begin{figure}[!htb]
\centering
\includegraphics[width=0.99\textwidth]{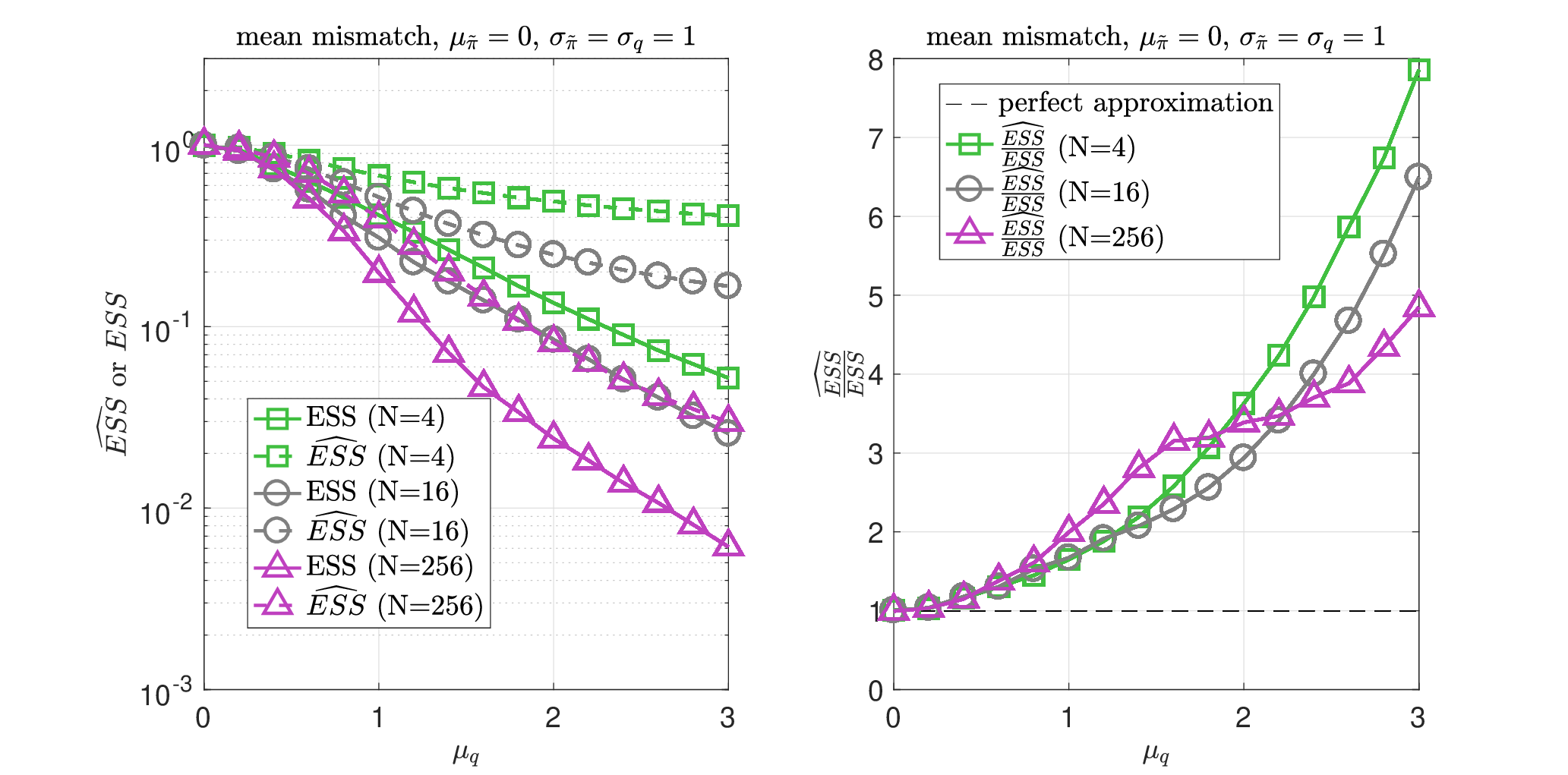}
\caption{Left plot shows the ESS and the $\ESS$ for the SNIS estimator of the mean of the target $\tilde \pi(x) = \mathcal{N}(x;0,1)$, when the mean of the proposal, $\mu_q$, is different from that of the target. The right plot shows the ratio between $\ESS$ and ESS. In all situations, the $\ESS$ overestimates ESS. Both  ESS and  $\ESS$ are normalized by the number of samples (divided by $N$), so they can be compared across different values of $N$.}
\label{toy_1_mu_mismatch}
\end{figure}

Let us consider the target $\normalized \pi(x) = \mathcal{N}(x;0,1)$. In Figure \ref{toy_1_mu_mismatch}, we show both the ESS and $\ESS$ of the SNIS estimator of the target mean, i.e. $h(x)=x$, with $N\in\{4,16,256\}$ samples, when $q(x) =\mathcal{N}(x;\mu_q,1)$, varying $\mu_q$ in the interval $[0,3]$. The values of ESS and $\ESS$ are normalized over $N$, so all cases can be compared. \cblue{We note that ESS and $\ESS$ are only available in the case without mismatch. In this example, we approximate both metrics through $10,000$ independent Monte Carlo runs.} In the right subplot, we display the ratio between $\ESS$ and ESS (i.e., when the line is above 1, the $\ESS$ overestimates the true ESS). Note that the efficiency deteriorates for large $N$. In all cases, $\ESS$ overestimates the ESS, with a larger gap when the mismatch between target and proposal increases. {In this example, all estimators have finite variance.}

\begin{figure}[!htb]
\centering
\includegraphics[width=0.99\textwidth]{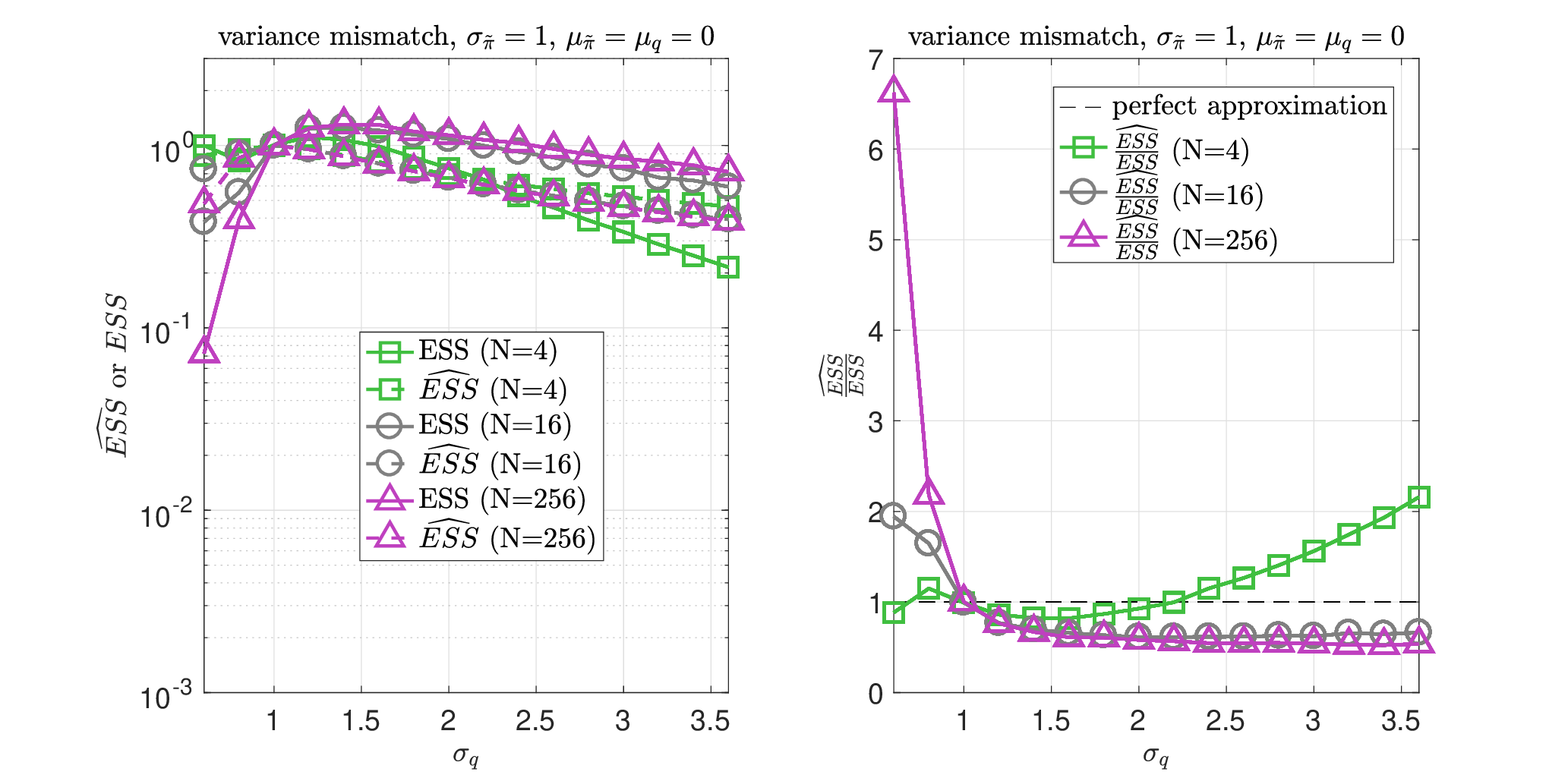}
\caption{Left plot shows the ESS and the $\ESS$ for the SNIS estimator of the mean of the target $\tilde \pi(x) = \mathcal{N}(x;0,1)$, when the variance of the proposal, $\sigma_q^2$, is different from that of the target. The right plot shows the ratio between $\ESS$ and ESS. Note that now the $\ESS$ sometimes underestimates and sometimes overestimates the true ESS. Both ESS and  $\ESS$ are normalized by the number of samples (divided by $N$), so they can be compared across different values of $N$.}
\label{toy_1_sig_mismatch}
\end{figure}

Figure \ref{toy_1_sig_mismatch} shows the results of the same experiment when there is a mismatch between the variances of target and proposal. In particular, the proposal is now $q(x) =\mathcal{N}(x;0,\sigma_q^2)$, and we vary $\sigma_q$ in the interval $[0.6,3.6]$ (note that too small a proposal variance can yield IS estimators with infinite variance). Note that again there is a gap between ESS and $\ESS$. Interestingly, for high values of $N$, and for high values of $\sigma_q$, the $\ESS$ now underestimates the ESS (contrary to the example with mean mismatch). This toy example contradicts the broad belief that states that if the $\ESS$ is \emph{high}, it is unclear if the quality of the approximation is good but if the $\ESS$ is \emph{low}, then necessarily the IS method is failing. One of the reasons for this contradiction is that in this example, the optimal proposal is broader than the target. 
As discussed above, the $\ESS$ is blind to $h(x)$ and also to cases where $ESS>N$. Note that in the left subplot of Fig. \ref{toy_1_sig_mismatch}, for $N\in\{16,256\}$ and in the range $\sigma_q\in[1,2.5]$, the ESS is larger than $N$ (larger than 1 in the normalized version displayed in the figure).\footnote {When $\sigma_q<1$, the variance of SNIS estimator can be infinite. This situation should be detected more easily with a different approach, while the $\ESS$ is clearly ineffective in detecting the malfunctioning.} 

\begin{figure}[!htb]
\centering
\includegraphics[width=0.99\textwidth]{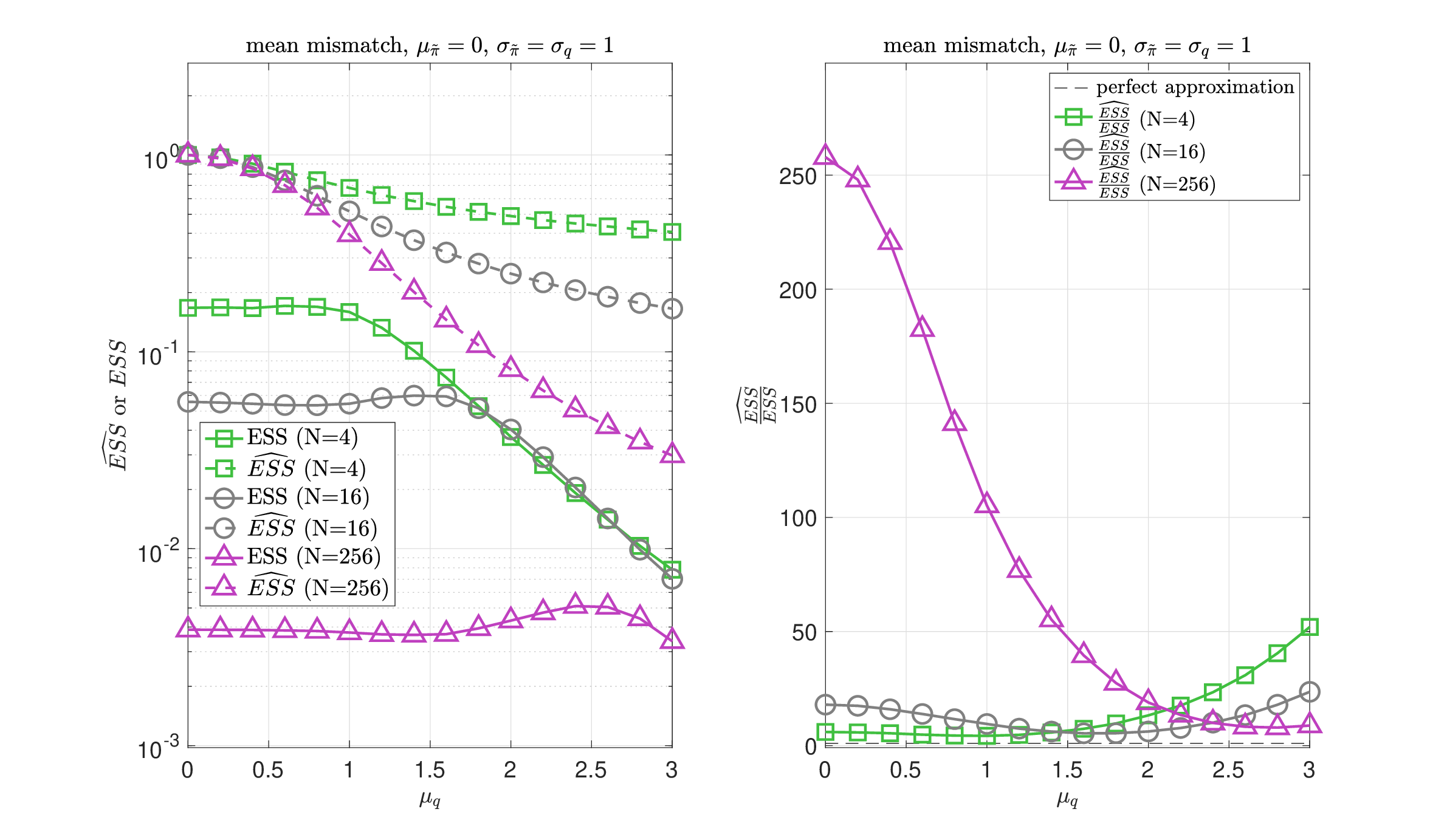}
\caption{\cblue{Left plot shows the ESS and the $\ESS$ for the SNIS estimator of the second moment of the target $\tilde \pi(x) = \mathcal{N}(x;0,1)$, when the mean of the proposal, $\mu_q$, is different from that of the target. The right plot shows the ratio between $\ESS$ and ESS. In all situations, the $\ESS$ overestimates ESS. Both  ESS and  $\ESS$ are normalized by the number of samples (divided by $N$), so they can be compared across different values of $N$.}}
\label{toy_1_mu_mismatch_x2}
\end{figure}

\begin{figure}[!htb]
\centering
\includegraphics[width=0.99\textwidth]{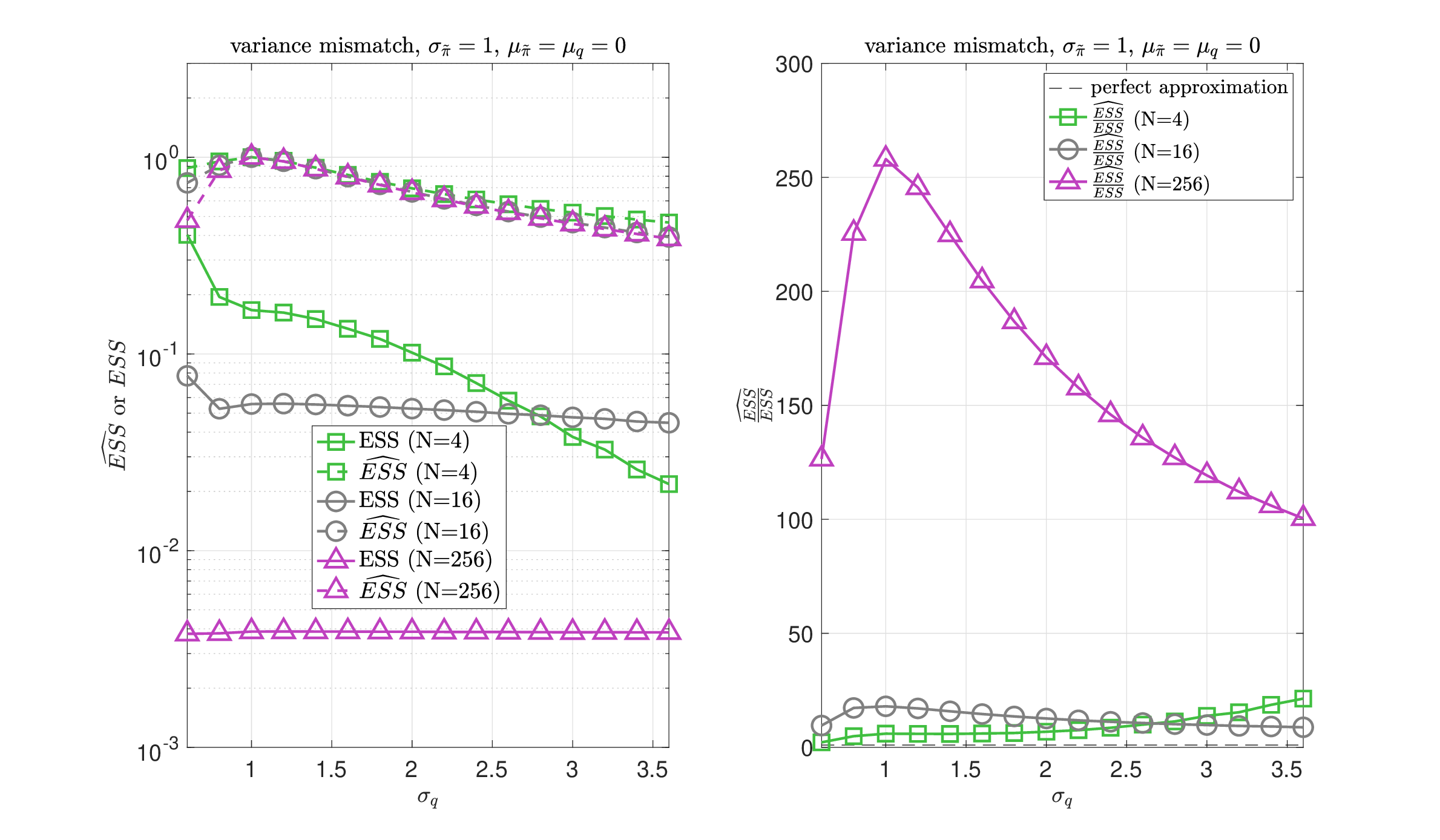}
\caption{\cblue{Left plot shows the ESS and the $\ESS$ for the SNIS estimator of the second moment of the target $\tilde \pi(x) = \mathcal{N}(x;0,1)$, when the variance of the proposal, $\sigma_q^2$, is different from that of the target. The right plot shows the ratio between $\ESS$ and ESS. Note that now the $\ESS$ sometimes underestimates and sometimes overestimates the true ESS. Both ESS and  $\ESS$ are normalized by the number of samples (divided by $N$), so they can be compared across different values of $N$.}}
\label{toy_1_sig_mismatch_x2}
\end{figure}

\cblue{We now repeat the experiment to estimate the second moment of the target $\tilde \pi(x)$, i.e., with $h(x) = x^2$. Figures \ref{toy_1_mu_mismatch_x2}  and \ref{toy_1_sig_mismatch_x2}  show, for the same setup, the ESS and $\ESS$ (left plot) and the the ratio between $\ESS$ and ESS (right plot). We note that the $\ESS$ is the same as in the previous setup, since the metric is not aware of $h(x)$, as thoroughly discussed in this paper. This explains why the mismatch between ESS and $\ESS$ is even more evident than in the previous case (with $h(x) = x$).}

\subsubsection{\cblue{Estimation of rare events}}
\label{sec_rare_toy}

\cblue{We now focus on a rare-event estimation problem \cite{bucklew2004introduction}. In particular, we consider a simplified version of the setup in \cite{owen2019importance,elvira2021multiple}, targeting the estimation of the integral in Eq. \eqref{eq_problem} with $h_{\alpha} = \mathbb{I}_{x \in \mathcal{S}_\alpha}(x)$, i.e., the indicator function that takes value one if $x\in\mathcal{S}_\alpha$ and zero otherwise. We consider the case with no mismatch between target and proposal, with $\tilde \pi(x) = q(x) = \mathcal{N}(0,\bI_{d_x})$, where $d_x$ is the dimension of $\x$, and $\bI_{d_x}$ is the identity matrix of such dimension. We choose $\mathcal{S}_\alpha = \Real^{d_x} \setminus [-\alpha, \alpha]^{d_x}$. For instance, for $d_x=1$, the region is $\mathcal{S}_\alpha = \{x: |x|>\alpha\}$. }

\cblue{This example is relevant because the efficiency of direct sampling decays when both $\alpha$ and $d_x$ grow. First, we note that, since function $h_{\alpha}$ is not present in $\ESS$, this metric cannot capture the efficiency loss that depends on $\alpha$. Moreover, in this case, we can also question the adequacy of the ESS as an efficiency metric per se, 
since regardless of $N$, all weights are always equal to $1$, and hence $\ESS=N$.} \cblue{The normalized weights are thus identical, and estimator $\widetilde I$ of Eq. \eqref{eq_SNIS}
 becomes the estimator $\bar I$ of Eq. \eqref{eq_estimator_target}. Hence, the original effective sample size of \eqref{eq_var_var} also yields $\text{ESS} = N$, regardless the choice of $h$.}

\begin{figure}[!htb]  
\centering
\includegraphics[width=0.69\textwidth]{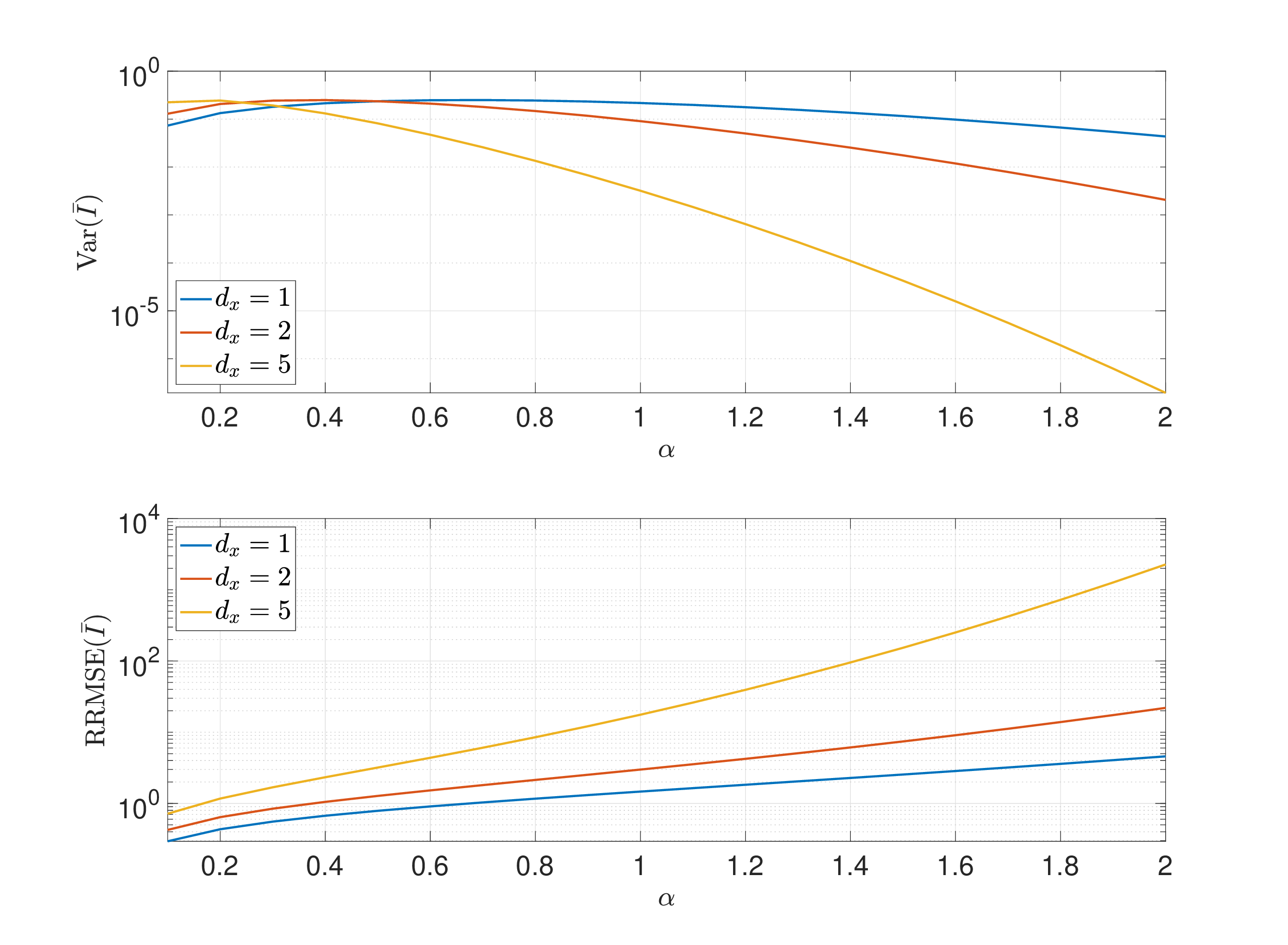}
\caption{Estimation of rare event with $\tilde\pi(x) = q(x) = \mathcal{N}(0,1)$, and \cblue{$h_{\alpha} = \mathbb{I}_{x \in \mathcal{S}_\alpha}(x)$, with $\mathcal{S}_\alpha = \Real^{d_x} \setminus [-\alpha, \alpha]^{d_x}$, with $d_x\in \{1,2,5\}$.} \textbf{Top:} Variance of standard Monte Carlo (and SNIS estimator) in log-scale when $\alpha$ grows. Note that for high values of $\alpha$, although the efficiency keeps deteriorating, the variance decays because the true value of the integral $I_\alpha$ becomes close to zero faster. \textbf{Bottom:} Relative root MSE (RRMSE) in log-scale of both estimators defined as $RRMSE = \frac{Var(\widetilde I)}{I_{\alpha} }$. \cblue{In this example, both ESS and $\ESS$ are equal to $N$ regardless $\alpha$ or $d_x$.}}
\label{rare_event_1}
\end{figure}

 \cblue{The top plot of Fig. \ref{rare_event_1} shows the variance of the SNIS estimator in log-scale, for $d_x \in [1, 2, 5]$ and with $N=10$ samples (we recall it is the same as the standard Monte Carlo estimator in this case). The bottom plot shows the relative root MSE (RRMSE) that normalizes the square root MSE by the true value of the integral $I_{\alpha}$.} \cblue{It can be seen that, when $\alpha$ grows, the efficiency decays: the RRMSE grows to infinity with $\alpha$, while $\text{ESS} = \ESS = N$ in the whole setup.}

\begin{figure}[!htb] 
\centering
\includegraphics[width=0.8\textwidth]{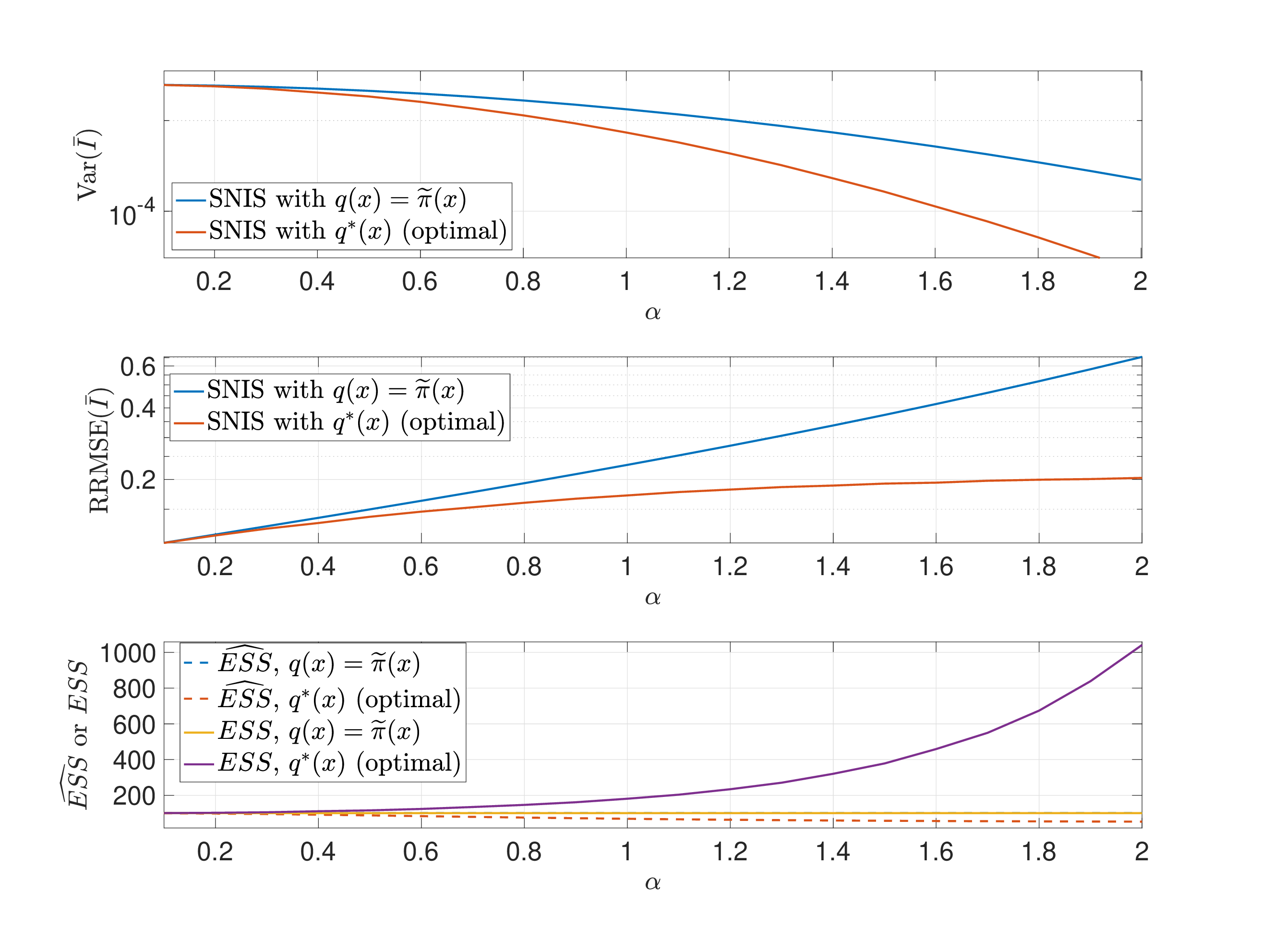}
\caption{\cblue{Estimation of rare event with SNIS estimators, with sub-optimal proposal $q(x) = \tilde\pi(x)  = \mathcal{N}(0,1)$ and optimal proposal $q^*(x)$.  The function is $h_{\alpha}(x) = \mathbb{I}_{x>\alpha}(x)$. 
\textbf{Top:} Variance of standard Monte Carlo (and SNIS estimator) in log-scale when $\alpha$ grows. Note that, for high values of $\alpha$, while the efficiency of SNIS keeps deteriorating, the variance decays because the true value of the integral $I_\alpha$ becomes close to zero faster. \textbf{Middle:} Relative root MSE (RRMSE) in log-scale of both estimators defined as $RRMSE = \frac{Var(\widetilde I)}{I_{\alpha} }$. \textbf{Bottom:} ESS and $\ESS$ for both SNIS estimators. The efficiency with the optimal proposal is superior to the case of the sub-optimal proposal, with an increasing gap when $\alpha$ grows. However, the averaged $\ESS$ is inferior in the optimal case, which again shows a misleading behavior of this metric.}}
\label{fig_rare_optimal}
\end{figure}

{We now consider instead the optimal SNIS proposal, $q^*(\x)\propto |h(\x) - I|  \pi(\x)$, and connect this experiment with \cite[Example 9.4]{owen2013montecarlo}. For simplicity, we particularize for $d_x = 1$ and $\mathcal{S}_\alpha = \{x: x>\alpha \}$. Since $0<I<1$ for any $\alpha>0$, it is easy to show that $q^*(\x) = \frac{1}{2I(1-I)}\left((1-I)\widetilde\pi(\x) \mathbb{I}_{\x \in \mathcal{S}_\alpha}(\x) + I\widetilde\pi(\x) \mathbb{I}_{\x \notin \mathcal{S}_\alpha}(\x)\right)$. Regardless $\alpha$, the optimal SNIS places half of the mass out of the region of interest $\mathcal{S}_\alpha$, which can be considered an inefficiency and explains why the UIS estimator is preferred for rare events estimation when it can be used (i.e., when $Z$ is known) \cite{owen2019importance,elvira2019multiple}. In Fig. \ref{fig_rare_optimal}, we show the variance (top) and the RRMSE (middle) of the SNIS estimator in two cases: with the sub-optimal proposal  $q(x) = \widetilde \pi(x)$ (as in the previous setup) and with the optimal proposal $q^*(x)$. Moreover, we display the ESS and $\ESS$ (bottom) for both estimators. The first remark is that, as expected, the optimal SNIS estimator obtains a better performance than the sub-optimal SNIS estimator. This is obviously translated into a higher ESS for the optimal SNIS. As discussed in Section \eqref{sec_conseq}, the ESS can be in general unbounded, and in this example, the ESS of the optimal SNIS estimator grows to infinity when $\alpha$ grows. However, the $\ESS$ associated to the optimal proposal does not grow (we recall that by construction is upper-bounded by $N$), and even more, it is inferior to the $\ESS$ of the sub-optimal SNIS. Moreover, the $\ESS$ of the optimal SNIS decays as $\alpha$ grows, while its ESS clearly grows.}

{As a summary, in this example the ESS metric and its approximation are unable to capture the loss of efficiency. This is because the ESS is a relative metric of the SNIS efficiency w.r.t. the standard Monte Carlo estimator, $\bar I$, which is considered as the gold standard in ESS, while it is known to be  highly inefficient in rare events \cite{owen2019importance}. Moreover, more sophisticated rare-event estimation methods based on IS \cite{owen2019importance,elvira2021multiple} can achieve an ESS of orders of magnitude above $N$, while the $\ESS$ will never be above $N$ by construction, as discussed in Section \ref{sec_conseq}. We strongly discourage the use of $\ESS$ in rare-events problems.}

\section{Beyond $\ESS$ for diagnostics in importance sampling}
\label{sec_beyond}

{In this section, we provide alternative derivations/interpretations of the $\ESS$ and we discuss its use when several proposals are used for simulating the samples.}

\subsection{Alternative derivations of $\ESS$}

The broad applicability of the $\ESS$ in the literature of particle-based methods has motivated alternative derivations or connections with other metrics. In the following, we present four of them.

\subsubsection{Relation of $\ESS$ with the discrepancy of the normalized weights}

\label{sec_euclidean_distance}

Let us consider the Euclidean distance $L_2$ between the uniform probability mass function (pmf) on $\mathcal{U}\{1,2,\ldots,N\}$, denoted as ${\overline \w}^* = [1/N,...,1/N]$, and the pmf described by ${\overline \w} = [\bar{w}_1,...,\bar{w}_N]$ on the same support: 
\begin{align}
L_2&=||{\overline \w}-{\overline \w}^*||_2 \nonumber \\
&= \sqrt{\sum_{n=1}^{N}\left(\bar{w}_n-\frac{1}{N}\right)^2}. \nonumber
\end{align}
Developing the square and re-arranging terms:
\begin{align}
L_2&=\sqrt{\left(\sum_{n=1}^{N}\bar{w}_n^2\right)+N\left(\frac{1}{N^2}\right)-\frac{2}{N} \sum_{n=1}^{N}\bar{w}_n} \nonumber\\
&=\sqrt{\left(\sum_{n=1}^{N}\bar{w}_n^2\right)-\frac{1}{N}} \nonumber \\
&=\sqrt{ \frac{1}{\ESS}-\frac{1}{N}}.
\end{align} 
{Alternatively, we can express this relation as
\begin{align} 
\ESS= \frac{1}{L_2^2 + \frac{1}{N}}.
\end{align} 
}
It is clear that decreasing the mismatch between the pmf described by the normalized weights, $\bar w_n$, and the uniform pmf is equivalent to increasing the $\ESS$. \cblue{Moreover, this connection allows to prove that $1\leq \ESS \leq N$ as follows. First, since $L_2\geq 0$, the maximum value of the approximated effective sample size is $\ESS=N$, when $L_2 =0$. Second, since ${\overline \w}$ is in the (N-1)-simplex, the vertices are the points that maximize the distance to the center, ${\overline \w}^*$. Their distance is $L_2 = \sqrt{1 - 1/N}$, which corresponds to the minimum approximated effective sample size,  $\ESS=1$, when ${\overline \w}$ is any permutation of the vector $[1, 0,...,0]$.} 
\cblue{This perspective of \emph{approximated} ESS as a discrepancy measure of the normalized weights, is exploited in \cite{huggins2015convergence,martino2017effective} to propose novel discrepancy measures with similar properties to $\ESS$ but with beneficial behavior in different situations. These alternative metrics are discussed in Section \ref{sec_alternative_metrics}.  
}

\subsubsection{$\ESS$ as loss of efficiency in a convex combination of independent r.v.'s}

In \cite[Section 9.4]{owen2013montecarlo}, the following interesting toy example is proposed. Let us consider $N$ i.i.d. r.v.'s $Z_n$, with $n=1,...,N$ with variance $\sigma_{Z}^2$ and mean $\mu_Z$. Let us assume a set of unnormalized weights $\textbf{w} = \{ w_n \}_{n=1}^N$ that are used to build the following linear combination
\begin{equation}
C_{\textbf{w}} = \frac { \sum _ { n = 1 } ^ { N } w _ { n } Z _ { n } } { \sum _ { n = 1 } ^ { N } w _ { n } }.
\end{equation}
Note that if all weights were the same, denoting $\textbf{w}^*$ as a vector with equal non-negative entries, the variance of $C_{\textbf{w}^*}$ would be $\sigma_{C_{\textbf{w}^*}}^2 = \frac{\sigma^2}{N}$. The question is, to how many $N_{\text{eff}}$ i.i.d. samples (instead of $N$), the weighted combination $C_{\textbf{w}}$ is equivalent to? The variance of $C_{\textbf{w}} $ can be readily computed as
\begin{align}
\sigma_{C_{\textbf{w}}}^2 &= \frac { \sum _ { n = 1 } ^ { N } w^2 _ { n } \sigma^2_Z } { \left( \sum _ { n = 1 } ^ { N } w _ { n } \right)^2}\nonumber \\
&= \frac{\sigma^2_Z} {\ESS},
\end{align}
i.e., the variance of the combination $C_{\textbf{w}}$ is equivalent to the variance of the equal-weights combination of $\ESS$ samples. Note that this example is far from being realistic in the intricate case of IS. In the case of single-proposal IS, the weights are also r.v.'s and not deterministic as in the previous example. Moreover, the weights are also dependent of the r.v.'s they are weighting, i.e., $W_n = \frac{\pi(\X_n)}{q(\X_n)}$ and $h(\X_n)$ are usually dependent r.v's. Finally, in the previous example, the variance of $Z_n$ is fixed, while in IS, depending if the samples are simulated from the proposal or the target, the r.v. $h(\X_n)$ has a different variance. However, this interesting and illustrative derivation highlights the lost of efficiency resulting from having a few weights dominate over the others.

\subsubsection{Relation of $\ESS$ with the coefficient of variation}
The derivation of \cite{Liu95} starts defining the coefficient of variation (CV) of the normalized weights as
\begin{equation} 
\CV = \sqrt{\frac{1}{N} \sum_{n=1}^N \left( N \bar w_n - 1 \right)^2}.
\end{equation}
Then, by basic manipulations, it can be shown that 
\begin{equation}
\ESS = \frac{N}{1+\CV^2} = \frac{1}{\sum_{n=1}^N \bar w_n^2}.
\end{equation}
The authors of \cite{Liu95} discuss that $\CV^2$ is a reasonable approximation of $\Var_q[W]$, assuming that $\E_q[W]=1$ (see the list of assumptions and approximations in Section \ref{sec_ass_and_app}). Note the relation of this derivation with the minimum square distance of Section \ref{sec_euclidean_distance}. In fact, the relation between CV and the Euclidean distance $L_2$ previously defined is $\CV = N \sqrt{L_2}$.

{ 
\subsubsection{$\ESS$ as approximation of the $\chi^2$ divergence between the target and proposal}  

\label{rho_interpr}

First, we show the connection between the second moment of the importance weights and the $\chi^2$ divergence between target and proposal (see for instance \cite{chen2005another}), that can be expressed as  
\begin{align}
\chi^2(\widetilde \pi, q) &= \int \frac{\left(\widetilde \pi(\x) - q(\x) \right)^2}{q(\x)}d\x \\
 &= \int \frac{ \widetilde \pi^2(\x)}{q(\x)}d\x - 1\\
 &= \E_q[W^2] - 1. 
\end{align}
The quantity $\rho = \E_q[W^2]$, which appears in the divergence above, has been studied for instance in \cite{agapiou2017importance} and \cite{whiteley2016role}, in both cases due to the connection to the $\chi^2(\tilde \pi, q)$ and for its relation with the $\ESS$.
\cblue{This quantity plays an important role since it helps to bound the error of the SNIS estimator under certain conditions \cite[Theorem 2.1]{agapiou2017importance}:
\begin{align}
\sup_{|h|\leq 1} \E \Big[ \left(\widetilde I - I \right)^2 \Big] \leq \frac{4}{N}\rho, \quad \forall N \geq 1.
\end{align}
By inspecting  Eq. \eqref{eq_ess_3a}, we note that $\ESS/N$ can be interpreted as a sample approximation of the quantity $\frac{\E_{q}[{W}]^2}{\E_{q}[{W}^2]}$. Therefore, for large enough $N$  we have that 
\begin{align}
\sup_{|h|\leq 1} \E \Big[ \left(\widetilde I - I \right)^2 \Big] \lesssim \frac{4}{\ESS}
\end{align}
as shown in \cite[Section 2.3.2]{agapiou2017importance}. 
Although the result is only asymptotic and limited to particular choices of $h$, it provides an interesting connection among the $\ESS$, the $\chi^2$ between target an proposal, and the efficiency of the SNIS estimator.} In relation with this result, \cite{whiteley2016role} show that by controlling the $\ESS$, it is possible to guarantee the long term stability of sequential Monte Carlo methods.
}
 
{
\subsection{Alternative metrics to $\ESS$}
\label{sec_alternative_metrics}

Due to the different drawbacks of  the $\ESS$  expression in Eq. \eqref{eq_rule_of_thumb} highlighted above,
several other approximations have been studied in literature          
\cite{Huggins15,martino2017effective}. In general, those metrics do not aim at improving the approximation of the ESS in \eqref{eq_var_var}, but instead look for alternative metrics with similar properties as the $\ESS$.

For instance, a metric called  perplexity, involving the discrete entropy \citep{Cover91} of the normalized weights was originally proposed in \citep{pmc-cappe08} (see also \cite[Chapter 4]{Robert10b} and \cite[Section 3.5]{Doucet08tut}). Another alternative metric that has shown a good performance in a variety of applications is defined as $\frac{1}{\max \left[\bar{w}_{1}, \ldots, \bar{w}_{N}\right]}$, i.e., the inverse of the maximum of the normalized weights ${\bar w}_n$ \citep{martino2017effective}. 
Here, we review a family of metrics, called \emph{Huggins-Roy's family}, introduced in \citep{Huggins15} and independently proposed, studied, and generalized in \citep{martino2017effective}. 
All these alternative metrics fulfill the required conditions described in the framework proposed in \citep{martino2017effective}. 
The Huggins-Roy's family  is defined as
$$
\begin{aligned}
\widehat{ESS}=H_{N}^{(\beta)}({\overline \w}) &=\left(\frac{1}{\sum_{n=1}^{N} \bar{w}_{n}^{\beta}}\right)^{\frac{1}{\beta-1}}, \\
&=\left(\sum_{n=1}^{N} \bar{w}_{n}^{\beta}\right)^{\frac{1}{1-\beta}}, \quad \beta \geq 0 .
\end{aligned}
$$
This family is related to the R\'enyi entropy of the probability mass function (pmf) defined by ${\{  \bar w}_n \}_{n=1}^N$ \citep{Cover91}. Indeed, the R\'enyi entropy is defined as
$$
R_{N}^{(\beta)}(\overline{\mathbf{w}})=\frac{1}{1-\beta} \log \left[\sum_{n=}^{N} \bar{w}_{n}^{\beta}\right], \quad \beta \geq 0
$$
Then, it is straightforward to note that
$$
H_{N}^{(\beta)}(\overline{\mathbf{w}})=\exp \left(R_{N}^{(\beta)}(\overline{\mathbf{w}})\right).
$$
Relevant specific cases are shown in Table \ref{TableLuca}.  However, more research is needed in order to  understand theoretically  these families of metrics, and to propose new alternatives to $\ESS$ that do not necessarily inherit some of its undesired properties.

\begin{table}[h]
\caption{Special cases of ESS measures contained in the Huggins-Roy's family $H_{N}^{(\beta)}(\overline{\mathbf{w}})$.}\label{TableLuca}
\small
{
\begin{tabular}{|c|c|c|c|c|}
 \hline
 $\beta=0$ & $\beta=1 / 2$ & $\beta=1$ & $\beta=2$ & $\beta=\infty$ \\
\hline \hline
 $||\overline\w ||_0$ & $\left(\sum_{n=1}^{N} \sqrt{\bar{w}_{n}}\right)^{2}$ & $\exp \left(-\sum_{n={w}}^{N} \bar{w}_{n} \log \bar{w}_{n}\right)$ & $\frac{1}{\sum_{n=1}^{N} \bar{w}_{n}^{2}}$ & $\frac{1}{\max \left[\bar{w}_{1}, \ldots, \bar{w}_{N}\right]}$ \\
$L_0$ norm of $\overline\w$ (number of non-zero elements) & & (perplexity) \cite{pmc-cappe08} & $\ESS$ \cite{Kong92} & \cite{martino2017effective} \\
\hline
\end{tabular}
}
\end{table}

}
\subsection{ ESS in Multiple Importance Sampling }
\label{sec_MIS}

As discussed in Section \ref{sec_ass_and_app}, the derivation of $\ESS$ assumes i.i.d. samples from a single proposal. 
However, $\ESS$ is used in situations where this is clearly not the case. The extension of single IS to multiple IS (MIS) is not unique, and many possible weighting and sampling schemes are possible (see a thorough review in \citep{elvira2019generalized}). This means that the $N$ samples and weights are r.v.'s that can follow a large set of possible distributions, and still build consistent estimators. Obviously, the assumptions and approximations from \eqref{eq_ess_mse} to \eqref{eq_rule_of_thumb} are in this case even more unrealistic than in single-proposal IS. For instance, if $N$ proposals are available and $N$ samples must be simulated, one can decide to deterministically sample once per proposal as $\x_n \sim q_n(\x)$, $n=1,...,N$, among many other options \citep[Section 3]{elvira2019generalized}. For this sampling scheme, several weighting schemes are also possible, e.g., the traditional interpretation of MIS (denoted as N1 in \citep{elvira2019generalized}) given by
\begin{equation}
  w_n = \frac{\pi(\x_n)}{q_n(\x_n)}, \qquad n=1,...,N.
\end{equation}
while the \emph{deterministic mixture} scheme (N3) implements the weights as 
\begin{equation}
  w_n = \frac{\pi(\x_n)}{\frac{1}{N} \sum_{j=1}^N q_j(\x_n)}, \qquad n=1,...,N.
\end{equation}
More details can be found in \citep{elvira2019generalized}, where it is shown that the UIS of N3 always present a better performance than N1 {in terms of variance of the UIS estimator}. The effect in the $\ESS$ of the different sampling and weighting schemes is yet to be understood as we show in the next example.

\subsubsection{\cblue{Toy example in MIS}}

Let us consider the multimodal target distribution,
$$\pi(x)=\frac{1}{3} \mathcal{N}(x;\nu_1,c^2) + \frac{1}{3} \mathcal{N}(x;\nu_2,c^2) + \frac{1}{3} \mathcal{N}(x;\nu_3,c^2),$$
with means $\nu_1=-3$, $\nu_2=0$, and $\nu_3=3$, and variance $c^2=1$. We implement MIS estimators with three proposals, $q_i(x)=\mathcal{N}(x;\mu_i,\sigma^2)$, $i=1,2,3$.

We select three scenarios depending on the mismatch between the target and the mixture of proposals:

\begin{enumerate}
  \item \textbf{Scenario 1: no mismatch.} The proposal means are $\mu_i = \nu_i$ {and} $i=1,2,3$  and the variance is $\sigma^2 = c^2 = 1$, i.e., the proposal pdfs can be seen as a whole mixture that exactly replicates the target, i.e., $\pi(x) = \psi(x) = \frac{1}{3} q_1(x) + \frac{1}{3} q_2(x) + \frac{1}{3} q_3(x)$. 
  \item \textbf{Scenario 2: mild mismatch.} The proposal means are $\mu_1 = -3$, $\mu_2 = -1$, and $\mu_3 = 3$, and the variance is $\sigma^2 = 2$.
 \item \textbf{Scenario 3: large mismatch.}   The proposal means are $\mu_1 = -4$, $\mu_2 = -1$, and $\mu_3 = 1$, and the variance is $\sigma^2 = 2$.
\end{enumerate}
In all situations, the goal is estimating the mean of the target pdf with the six MIS schemes proposed in \citep[Section 5]{elvira2019generalized}. The schemes N1 and N3 are described above, and the R3 index refers to the schemes where samples are simulated from the mixture of proposals, $\psi(x)$, and all the weights are computed as $w_n = \frac{\pi(x)}{\psi(x)}$. 

\begin{figure}[!htb]
\centering
\includegraphics[width=0.99\textwidth]{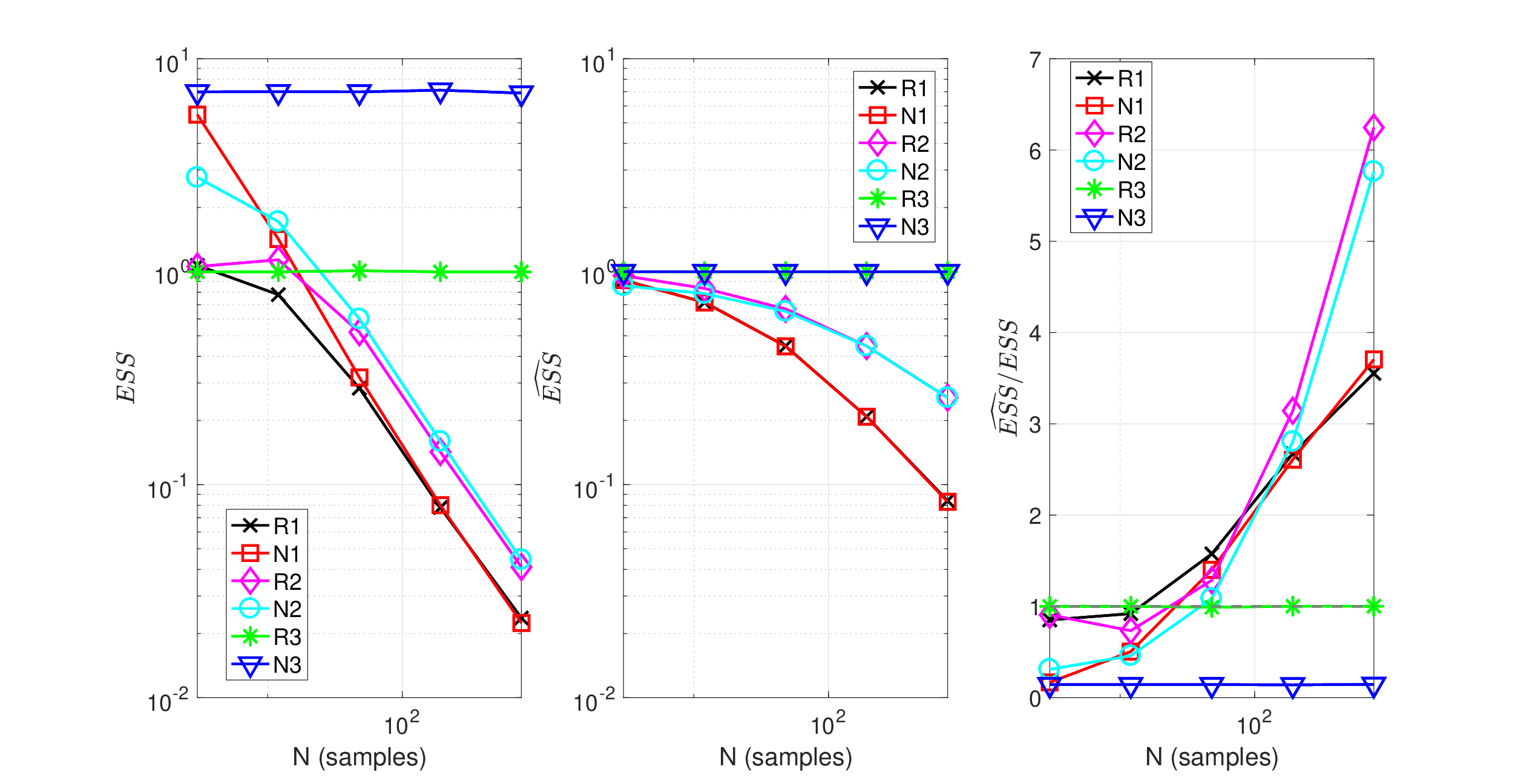}
\caption{{\textbf{MIS example (Scenario 1).} True ESS (left), $\widehat {ESS}$ (middle), and ratio of both (right) of the self-normalized IS estimator of the target mean, for the different MIS schemes as the total number of samples $N$ grows. Note that {ESS} is normalized, i.e. divided by the number of samples $N$.}}
\label{MIS_no_mismatch}
\end{figure}

Figure \ref{MIS_no_mismatch} shows, for the case of no mismatch (Scenario 1), the ESS, the $\ESS$, and the ratio of both quantities, for all MIS schemes and for different values of total number of samples $N$ (from $3$ to $3\times2^9$). Both ESS and $\ESS$ are normalized (divided by $N$) so the behavior of the later can be analyzed. First, note that in this example, the scheme R3 corresponds to direct sampling from the target distribution and its ESS is $1$ regardless of the number of total samples drawn, which explains why it is the only scenario and scheme where $\ESS$ is accurate. Note that the $\ESS$ of N3 is around $7$ times smaller than the true ESS. The reason is that the N3 proposal in this scenario samples from the target with variance reduction (there are the same number of samples for each component of the mixture, rather than a random number as in an i.i.d.~simulation from the mixture). Note that for most schemes, $\ESS$ overestimates the ESS, when $N$ grows, but this is not the case for low values of $N$. 

It is very illustrative that $\ESS$ is blind to the difference between $N3$ and $R3$. In both schemes, due to the perfect match between the target and the mixture of proposals, all weights are always equal to 1, and hence the $\ESS$ is maximum. However, $N3$ simulates the samples with variance reduction, which cannot be captured by $\ESS$ since the samples play no role in its calculation.

\begin{figure}[!htb]
\centering
\includegraphics[width=0.99\textwidth]{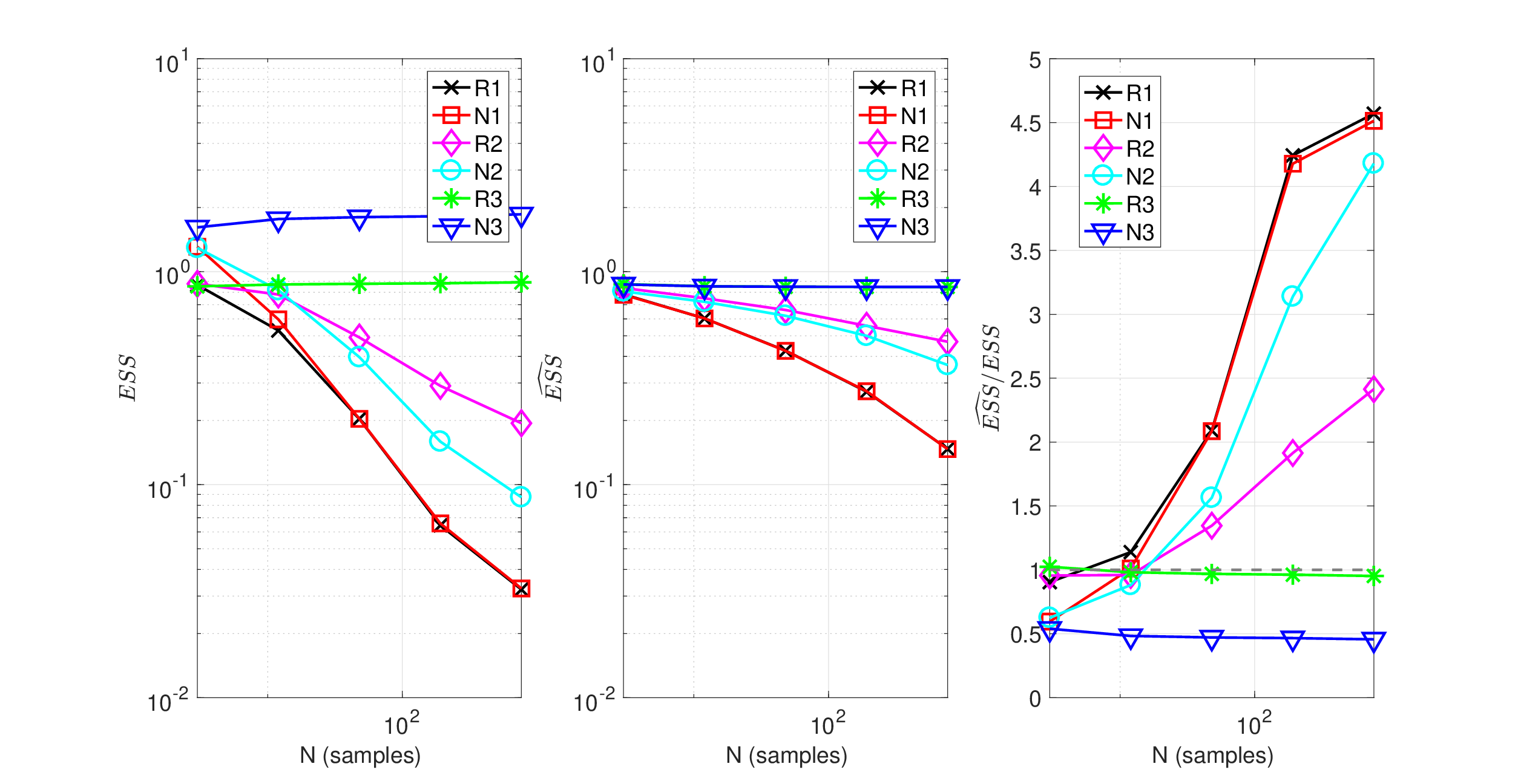}
\caption{{\textbf{MIS example (Scenario 2).} True ESS (left), $\widehat {ESS}$ (middle), and ratio of both (right) of the self-normalized IS estimator of the target mean, for the different MIS schemes as the total number of samples $N$ grows. Note that {ESS} is normalized, i.e. divided by the number of samples $N$.}}
\label{MIS_mild_mismatch}
\end{figure}

\begin{figure}[!htb]
\centering
\includegraphics[width=0.99\textwidth]{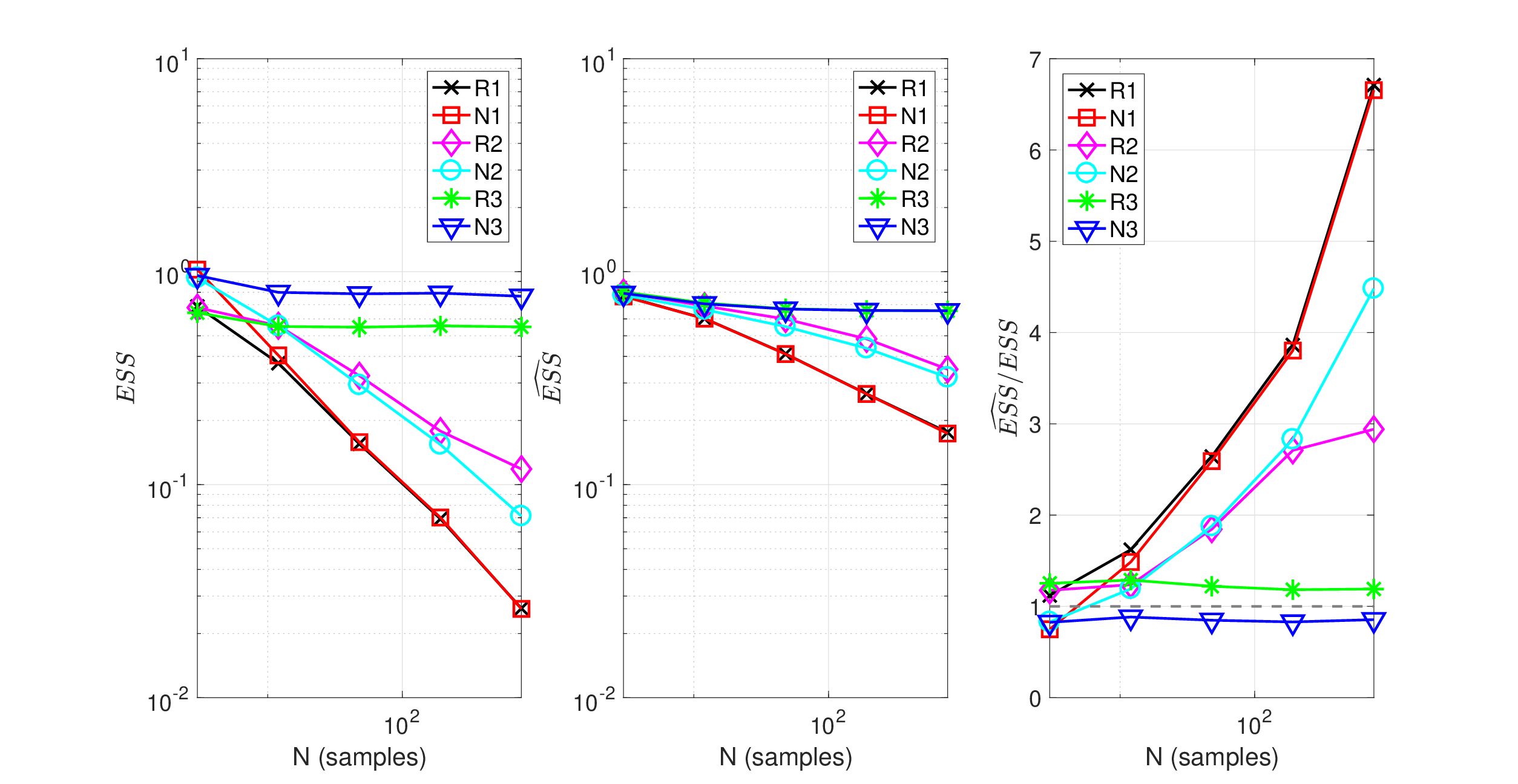}
\caption{{\textbf{MIS example (Scenario 3).} True ESS (left), $\widehat {ESS}$ (middle), and ratio of both (right) of the self-normalized IS estimator of the target mean, for the different MIS schemes as the total number of samples $N$ grows. Note that {ESS} is normalized, i.e. divided by the number of samples $N$.}}
\label{MIS_large_mismatch}
\end{figure}

Figure \ref{MIS_mild_mismatch} shows the same curves for the case of mild mismatch between the target and the mixture of proposals (Scenario 2). In this scenario, all MIS schemes loose efficiency due to the mismatch, but $\ESS$ keeps underestimating the ESS in $N3$.
Finally, Figure \ref{MIS_large_mismatch} shows the same curves when the mismatch between the target and the mixture of proposals is large (Scenario 3). In this scenario, the $\ESS$ underestimates the ESS in all MIS schemes except for $N3$.

\subsubsection{An alternative $\ESS$ in MIS}

An appropriate extension of Eq. \eqref{eq_ess_1} to MIS is not straightforward, since the samples are drawn from different proposal pdfs. Noting that in standard IS  $\Var_q[\widehat{Z}]=\frac{\Var_q[w]}{N}$, we propose the following natural extension of the ESS to the MIS approach:
\begin{equation}
ESS_{\text{MIS}}  \approx \frac{N}{1+ \Var [\widehat{Z}]},
\label{eq_ess_mis_1}
\end{equation} 
where the variance of $\widehat Z$ is now calculated taking into account the whole set of proposals. In this case, $\Var [\widehat{Z}]$ needs to be still estimated. Note also that the approximation from the true ESS to Eq. \eqref{eq_ess_1} assumes a single proposal (and a normalized target).

{
\section{{New directions for alternative metrics}}
\label{sec_new}

In Section \ref{sec_alternative_metrics}, we have revisited some of the alternative metrics to the  $\ESS$ that have been recently proposed.
However, all those metrics inherit some of the undesired properties of the $\ESS$, as we discussed in Section \ref{sec_conseq}. 
We now explore two research lines that aim at dropping two of those properties, namely the lack of presence of both the function $h$ and the set of samples $\left\{\mathbf{x}_{n}\right\}_{n=1}^{N}$.
 
}
\subsection{\cblue{$h(\x)$-aware $\ESS$}}

\cblue{In most estimation problems, the function $h(\x)$ plays a crucial role in the efficiency of the estimators, e.g., in the rare-event estimation problem of Section \ref{sec_rare_toy}.}  Note that, for the optimal {UIS} proposal $q^*(\x)\propto |h(\x)|\tilde \pi(\x)$, all evaluations $\frac{|h(\x_n)| \tilde \pi(\x_n)}{q^*(\x_n)}$ are identical regardless of the value of $\x_n$. Hence, a natural extension of the $\ESS$ is considering
\begin{equation}
\ESS^{(h)} = \frac{1}{\sum_{n=1}^N \left( \bar w_n^{(h)} \right)^2}
\end{equation}
where
\begin{equation}
\bar w_n^{(h)} = \frac{|h(\x_n)|w_n}{\sum_{j=1}^N |h(\x_j)|w_j}.
\end{equation}
\cblue{In this case, if the samples are simulated from the optimal UIS proposal $q^*(\x)$, then $\ESS^{(h)} = N$ always holds for non-negative functions $h$ (i.e., it takes the maximum value). It is important to note that this modification goes beyond the definition of Eq. \eqref{eq_var_var} (or the proposed variation with the MSE in the denominator in Eq. \eqref{eq_ess_mse}). The original definition implicitly assumes that simulating from the target $\normalized \pi(\x)$ is the benchmark to compare with. However, it is well-known that this is not the case when attempting to reduce the variance of the IS estimator, which is clearly dependent of $h$.}

\cblue{Other alternative approximations could be considered in the derivation of Section \ref{sec_derivation}. For instance, the natural variance estimate of SNIS estimator, proposed in \cite[Sec. 9.1]{owen2013montecarlo} and recently used in \cite{nilakanta2020output}, could be plugged in the derivation. The difficulty of this approach is that the original derivation approximates $\Var_q[\widetilde I]$ in such a way it depends on $\Var_{\tilde \pi}[\bar I]$, and therefore this term cancels out in the ratio of Eq. \eqref{eq_ess_1}. This is particularly convenient, since $\Var_{\tilde \pi}[\bar I]$ is arguably the hardest term to estimate.}

\subsection{\cblue{Sample-aware $\ESS$}}
 
The diversity of the samples is known to increase the performance of the estimators \cite[Chapter 8]{owen2013montecarlo}. The diversity in the set of samples is also key in IS-based methods that incorporate resampling steps (e.g., particle filtering \citep{li2015resampling} or adaptive IS \cite{bugallo2017adaptive}) to avoid degeneracy, even in the case where the $\ESS$ is high. 
For these reasons, we would like the set of samples $\{\x_n \}_{n=1}^N$, to play a role in the approximated ESS. Intuitively, when the samples are farther apart this should yield larger ESS (for a fixed set of weights). An approximated ESS that would remain constant (in average) after a resampling step would be also desirable in order to avoid the well-known paradox that $\ESS = N$ after a resampling step.

Here we follow the ideas of partitioning the space suggested in recent works in the literature \cite{Li12,GIS}, which have shown to improve the quality of the resampling steps in sequential Monte Carlo schemes \cite{CMC}. The main idea is to divide the domain $\mathcal{X}$ of the random variable $\mathbf{X}$ into $M$ separate mutually exclusive regions. More specifically, let us consider an integer $M \in \mathbb{N}^{+}$, and a partition $\mathcal{P}=\left\{\mathcal{X}_{1}, \mathcal{X}_{2},...,\mathcal{X}_{m}\right\}$ of the state space with $M$ disjoint subsets,
$$
\begin{aligned}
&\mathcal{X}_{1} \cup \mathcal{X}_{2} \cup \ldots \cup \mathcal{X}_{M}=\mathcal{X}, \\
&\mathcal{X}_{i} \cap \mathcal{X}_{k}=\emptyset, \quad i \neq k, \quad \forall i, j \in\{1, \ldots, M\} .
\end{aligned}
$$
Let consider $N$ weighted samples $\left\{\mathbf{x}_{n},w_{n}\right\}_{n=1}^{N}$
such that $N>>M$.
We denote the subset of the set of indices $\{1, \ldots, N\}$,
$$
\mathcal{J}_{m}=\left\{i=1, \ldots, N: \x_{i} \in \mathcal{X}_{m}\right\},
$$
which are associated with the samples in the $m$-th sub-region $\mathcal{X}_{m}$. The cardinality $\left|\mathcal{J}_{m}\right|$ denotes the number of samples in $\mathcal{X}_{m}$, and we have $\sum_{m=1}^{M}\left|\mathcal{J}_{m}\right|=N$. We can associate a {\it summary weight} for each region $\mathcal{X}_{m}$, as
\begin{eqnarray}
\bar{\nu}_m=\sum_{i \in \mathcal{J}_{m}} \bar{w}_{i}.
\end{eqnarray}
Note that $\sum_{m=1}^M \bar{\nu}_m=1$, since $\bar{w}_{n}=\frac{w_{n}}{\sum_{i=1}^N w_{i}}$.
We define a summary metric, alternatively to $\ESS$, as
\begin{eqnarray}
\widehat{ESS}^{(x)}=\frac{1}{\sum_{m=1}^M \bar{\nu}_m^2},  \qquad  1\leq \widehat{ESS}^{(x)}\leq M,
\end{eqnarray}
that measures the dissimilarity between the summary weight of each partition. 
We can also define new re-normalized weights (to sum up to one in each subset) as
\begin{eqnarray}
\bar{w}_{n,m}=\frac{w_n}{\sum_{i \in \mathcal{J}_{m}} \bar{w}_{i}}=\frac{w_n}{\bar{\nu}_m},
\end{eqnarray}
and  define a \emph{local} $\ESS$ of each subset as
\begin{eqnarray}
\widehat{ESS}_m=\frac{1}{\sum_{n \in \mathcal{J}_{m}} \bar{w}_{n,m}^2}, \qquad  1\leq \widehat{ESS}_m\leq |\mathcal{J}_{m}|.
\end{eqnarray}
It is easy to show that $M \leq \sum_{m=1}^M \widehat{ESS}_m \leq N$.

It is possible to find the standard $\ESS$ as a function of the $M+1$ metrics defined above (i.e., the $M$ local metrics and the $\widehat{ESS}^{(x)}$) as: 
\begin{eqnarray}
\widehat{ESS}=\widehat{ESS}^{(x)} \left(\frac{1}{M}\sum_{m=1}^M \widehat{ESS}_m\right), \qquad  1\leq\widehat{ESS}\leq N.
\label{eq_connection}
\end{eqnarray}
We note that $\widehat{ESS}^{(x)}$ penalizes the concentration of the samples in the space. The $M+1$ metrics give then more information about the diversity in the space of the set of samples, while the $\ESS$ is again recovered by compressing this information as in Eq. \eqref{eq_connection}.    

This approach raises interesting questions for future research. For instance, it is possible to design the optimal choice of the partition. It is possible to study the behavior of $\widehat{ESS}$ if the partition is done empirically, e.g., with a clustering algorithm. Instead of using the same structure as in $\ESS$,  we could use other metrics in the literature (e.g.,  those revisited in Section \ref{sec_alternative_metrics}) to build $\widehat{ESS}^{(x)}$  and the set of local metrics $\{ \ESS_m \}_{m=1}^M$.  

Finally, there is a connection of these sample-aware metrics with the metric $\ESS$-IGH, proposed in \cite[Eq. (23)]{elvira2020importance} in the context novel quadrature methods derived under an importance sampling perspective. This metric, can be seen as a generalization of $\ESS$ and penalizes situations when large weights correspond to samples that are farther apart from the mean of the proposal, i.e., it incorporates information from the sample space.

{
\section{{Conclusion}}
\label{sec_conclusions}

Since the $\ESS$ metric was derived in \cite{Kong92}, it has been widely used in the literature of importance sampling. In this paper, we have revisited its derivation, highlighting the assumptions and approximations, and summarizing the undesired properties. Moreover, we have shown in representative numerical examples that the metric can produce misleading conclusions. The purpose of this paper is to raise awareness about the $\ESS$, so it can be applied carefully.  

We have also described alternative derivations and interesting connections with other metrics. We have revisited alternative metrics that have been recently proposed in the literature, and we have discussed potential improvements of the metric. Namely, we have discussed the generalization for the case of multiple proposals, the incorporation of the function $h(\x)$ which is key in the efficiency of the SNIS estimator, and the inclusion of information that is present in the set of samples. Our preliminary exploration will allow for further research in those directions.

In the future, we would like to use the unnormalized IS weights to evaluate a sequence of particle approximations. Even if they do not have an absolute measure of performance (because $Z$ is unknown), in adaptive IS schemes, one could exploit this information in iterative schemes. 
Finally, it would be desirable to have a more stable particle approximation. Note that the unnormalized IS weights often follow a right skewed distribution, which translates into having many zeros in the normalized weights. We could either use the \emph{median} instead of the mean. We could also nonlinearly transform the unnormalized weights before normalizing them as in \cite{ionides2008,koblents2013population,vehtari2015pareto} (see a review in \cite{martino2018comparison}).
}

\section{Acknowledgments}

The authors want to thank Matthew Ludkin and the referees for their useful comments.
 
\appendix

\section{Delta method for the approximation of expectations}
\label{appendix_delta_1}
 
\cblue{In this appendix, we provide a generic derivation of the delta method, which uses a second order Taylor expansion to approximate expectations of functions that involve dependent r.v.'s. In the main text, we particularize twice this generic formulation, making explicit the two delta methods involved in the derivation of $\ESS$.}
 
Let us define a bivariate function $f(T,Y)$. A second order expansion around a point $(T_0,Y_0)$ is 
\begin{align}
f(T,Y) &\approx f(T_0,Y_0) + f_T'(T_0,Y_0)\left(T-T_0\right) + f_Y'(T_0,Y_0)\left(Y-Y_0\right)\nonumber\\
&+\frac{1}{2}\left(f_{TT}''(T_0,Y_0)\left(T-T_0\right)^2  +2f_{TY}''(T_0,Y_0)\left(T-T_0\right)\left(Y-Y_0\right) + f_{YY}''(T_0,Y_0)\left(Y-Y_0\right)^2 \right).
\label{taylor_order_2}
\end{align} 
Therefore, a good approximation to $\E[f(T,Y)]$ with $f(T,Y)$ expanded around $(T_0,Y_0)$ is 
\begin{align}
\E[f(T,Y)] &\approx f(T_0,Y_0) +\frac{1}{2}\left(f_{TT}''(T_0,Y_0)\Var[T] +2f_{TY}''(T_0,Y_0)\Cov[T,Y] + f_{YY}''(T_0,Y_0)\Var[Y] \right).
\label{delta_expectation}
\end{align}
Now we derive an approximation of the variance of $f(T,Y)$, that by definition, is expressed as
\begin{align}
\Var[f(T,Y)] = \E\Big[\left( f(T,Y)  - \E[ f(T,Y) ] \right)^2\Big].
\label{delta_var_1}
\end{align}
We take a rough approximation $\E[f(T,Y)] \approx  f(T_0,Y_0) $ from Eq. \eqref{taylor_order_2}, as it is done for instance in \cite{seltman2012approximations}. Then, by taking a first order expansion $f(T,Y) \approx f(T_0,Y_0) + f_T'(T_0,Y_0)\left(T-T_0\right) +  f_Y'(T_0,Y_0)\left(Y-Y_0\right)$, Eq. \eqref{delta_var_1} turns into
\begin{align}
\Var[f(T,Y)] &\approx  \E\Big[\left( f(T_0,Y_0) + f_T'(T_0,Y_0)\left(T-T_0\right) + f_Y'(T_0,Y_0)\left(Y-Y_0\right)  - f(T_0,Y_0) ) ] \right)^2\Big] \nonumber\\
&= \E\Big[\left(  f_T'(T_0,Y_0)\left(T-T_0\right) + f_Y'(T_0,Y_0)\left(Y-Y_0\right)  ) ] \right)^2\Big] \nonumber\\
&= \E\Big[  \left( f_T'(T_0,Y_0)\left(T-T_0\right)\right)^2 + \left(f_Y'(T_0,Y_0)\left(Y-Y_0\right)\right)^2  \nonumber \\
& \;\;\; + 2f_T'(T_0,Y_0)\left(T-T_0\right)f_Y'(T_0,Y_0)\left(Y-Y_0\right)\Big] \nonumber\\
&= f_T'^2(T_0,Y_0) \Var[T] + f_Y'^2(T_0,Y_0) \Var[Y] + 2f_T'(T_0,Y_0) f_Y'(T_0,Y_0) \Cov[T,Y].
\label{delta_var_2}
\end{align}





\selectlanguage{english}
\bibliographystyle{Chicago}

\end{document}